# Computational Discovery of Fast Interstitial Oxygen Conductors

Jun Meng[1,4], Md Sariful Sheikh[1,4], Ryan Jacobs[1], Jian Liu[2], William O. Nachlas[3], Xiangguo Li[1], Dane Morgan[1,*]

*[1] Department of Materials Science and Engineering, University of Wisconsin Madison, Madison, WI, USA.*
*[2] DOE National Energy Technology Laboratory, Morgantown, WV, USA.*
*[3] Department of Geoscience, University of Wisconsin Madison, Madison, WI, USA.*
*[4] These authors contributed equally: Jun Meng, Md Sariful Sheikh. E-mails: ddmorgan@wisc.edu.*

New highly oxygen-active materials may enhance many energy-related technologies by enabling efficient oxygen-ion transport at lower temperatures, e.g., below ≈400 ˚C. Interstitial oxygen conductors have the potential to realize such performance but have received far less attention than vacancy-mediated conductors. Here, we combine physically-motivated structure and property descriptors, *ab initio* simulations, and experiments to demonstrate an approach to discover new fast interstitial oxygen conductors. Multiple new families were found which adopt completely different structures from known oxygen conductors. From these families, we synthesized and studied oxygen kinetics in $La_4Mn_5Si_4O_{22+\delta}$ (LMS), a representative member of perrierite/chevkinite family. We found LMS has higher oxygen ionic conductivity than the widely used yttria-stabilized $ZrO_2$, and among the highest surface oxygen exchange rates at intermediate temperature of known materials. The fast oxygen kinetics is the result of simultaneously active interstitial and interstitialcy diffusion pathways. This work developed and demonstrated a powerful approach for discovering new families of interstitial oxygen conductors and suggests many more such materials remain to be discovered.

Materials which rapidly conduct oxygen are critical for a variety of energy devices such as fuel cells (solid oxide,[1,2] proton ceramic,[3,4] and reversible[5] cells), electrolyzers,[6] solid-oxide metal-air redox batteries,[7] gas sensors,[8] chemical looping devices,[9] memristors,[10] and oxygen separation membranes.[11] Almost all of the state-of-the-art oxygen-active materials transport oxygen via a vacancy-mediated mechanism, which requires mobile oxygen to move off the oxygen sublattice and pass through an unstable activated state. The bond breaking associated with this process is generally energetically unfavorable and, consequently, even the best vacancy oxygen conductors have inadequate oxygen kinetics for practical applications at temperatures below ≈ 600 ˚C. Overcoming this limitation would allow for more varied, durable, and cost-effective devices. The vacancy-mediated mechanism has dominated the science of oxygen-active materials since the discovery of the first oxygen ion conductor (Y doped $ZrO_2$) around 1900 by Nernst [12] and continues to present. In contrast, interstitial oxygen diffusion is relatively uncommon and there are no systematic approaches to discover or optimize interstitial oxygen conductors.

Interstitial oxygen conductors have many potential advantages over vacancy-mediated conductors. Interstitial oxygen diffuses within the interstitial lattice, therefore the interstitial oxygen conductors typically have lower migration barriers compared to vacancy oxygen conductors, with entries in the Citrine Informatics database showing average values of ≈0.6 eV versus ≈1.1 eV, respectively (**Fig. 1**).[13] This ≈0.5 eV reduction in migration barrier would afford an approximate ×1000 increase in ionic conductivity at 600 °C. Another benefit is that, at fixed $P(O_2)$, interstitial oxygen becomes thermodynamically more favorable as the temperature is decreased (i.e., more oxidizing conditions). This higher defect concentration will increase the oxygen conductivity at low temperatures, opposite the trend in vacancy conductors. Other potential benefits may include a trend of increasing vs. decreasing diffusivity at higher defect concentration, since at least one study[14] found that interstitial oxygen diffusivity in $CeO_2$ increased beyond the dilute limit while vacancy oxygen diffusivity decreased. Finally, oxygen surface exchange and catalytic reactions involving pulling oxygen to the surface are potentially faster when transport is mediated by interstitials as compared to vacancies, as the entire surface has accessible interstitial sites to absorb oxygen as interstitials. The synergistic combination of lower migration barriers, increasing concentration at low temperatures, and many active surface sites for exchange suggest that interstitial oxygen conductors may lead to large performance improvements in oxygen-active materials. Therefore, we propose that a promising path to expand the palette of highly oxygen-active materials at lower temperatures is to develop methods to discover and engineer interstitial oxygen conductors.

The imbalance in the number of known vacancy versus interstitial oxygen conductors is likely the result of the difficulty of forming interstitial oxygen in many materials, due to the large size of the oxygen anion. The oxides presently known to have predominantly interstitial-mediated oxygen conductivity remain constrained to five families, which include Ruddlesden-Popper (e.g., $La_2NiO_{4+\delta}$ [15]), apatite (e.g. $La_{10-x}Sr_xSi_6O_{27-0.5x}$[16]), melilite (e.g., $La_{2-x}Sr_xGa_3O_{7+0.5x}$[17]), hexagonal manganites (e.g., $YMnO_{3+\delta}$[18]), and hexagonal perovskite (e.g., $Ba_7Nb_{3.9}Mo_{1.1}O_{20.05}$[19]). In addition to these purely interstitial-conducting materials, the fluorite-type (e.g., $UO_{2+\delta}$[20]), and scheelite (e.g., $CeNbO_{4+\delta}$[21,22]) compounds can conduct oxygen ions through both interstitial- and vacancy-mediated mechanisms, depending on the impurity type. A few of these systems (Ruddlesden-Popper $Nd_2NiO_{4+\delta}$,[23] apatite $La_{9.75}Sr_{0.25}Si_6O_{26.895}$,[16] melilite $La_{1.54}Sr_{0.46}Ga_3O_{7.27}$,[24] and hexagonal perovskite $Ba_7Nb_{3.9}Mo_{1.1}O_{20.05}$[25]) were reported to show high ionic conductivity, comparable to the commercial ionic conductor yttria-stabilized zirconia (YSZ). Despite this promising list, far less attention has been devoted to interstitial-dominated systems than vacancy-dominated ones, and we lack methods to discover new interstitial oxygen conductors. In this work we overcame that limitation and laid a foundation to dramatically increase the palette of interstitial oxygen systems, enabling researchers to explore the advantages of interstitial conductors over their more conventional vacancy-mediated counterparts for the advancement of oxygen-active material applications.

In this work, we proposed a practical approach based on structural and chemical features as well as *ab initio* calculations for finding high-performing interstitial oxygen conductors, identified multiple new promising classes, and demonstrated the exceptional performance of one example material, $La_4Mn_5Si_4O_{22+\delta}$ (LMS). To the best of our knowledge, LMS has no similarity to any known oxygen ion conductors and cannot be related to known oxygen active materials by simple substitutions, structural similarity arguments, or other approaches that might render it “obvious” in some way. The discovery of LMS demonstrated the effectiveness of the approach to discover new material classes that would likely not have been considered for oxygen-active applications without such computational guidance.

## A descriptor approach for discovering new interstitial oxygen conductors

We proposed a set of simple descriptors (features) for discovering interstitial oxygen conductors based on material structure, composition, and readily available property data. The structure features are based on the hypothesis that (1) the facile formation of interstitial oxygen requires sufficient free volume and electrons from oxidizable transition metal cations, (2) the fast migration of interstitial oxygen should be enhanced by the presence of short diffusion pathways, a result consistent with intuition and the correlation between hop length and migration barrier found in perovskites.[26] The property features are focused on thermodynamic stability and synthesizability. This led to screening on the following five structure and property features: (1) free space for interstitial oxygen, (2) short hop distance, (3) thermodynamic stability, (4) oxidizability, and (5) synthesizability. One could quantify these features in different ways and here we use structures and properties from the Materials Project[27], with the details given in **Methods 1,** step (1)-(5). We screened nearly 34k oxide materials with the five descriptors and retained 519 compounds, which were classified into 345 unique structural groups based on structural similarity analysis[28] (**Methods 1,** step (6)). One candidate for each group was selected for further validation with *ab initio* studies, discussed below. It is striking to note the power of these simple descriptors based on physical-intuition-guided hypotheses, as we quickly winnowed the field of 34k oxides down to 345, a 99% reduction in the search space enabled by basic analysis of material structure and composition.

## *Ab initio* computational screening of promising compounds for discovering new interstitial oxygen conductors

The above 6 screening steps based on structure, composition, and property data took only a couple of days on a fast processor. Then we screened the resulting 345 compounds for promising cases by the formation energy ($E_f$) and migration barriers ($E_m$) of interstitial oxygen, calculated by the slower but more quantitative density functional theory (DFT) methods ("*ab initio* simulation" stage in **Fig. 2**). Specifically, compounds with $E_f \leq 0.3$ eV (at P($O_2$) = 0.2 atm and T=300 K) were considered promising (**Methods 2,** step (7)), and 80 out of the 345 compounds (23%) were retained (**Table S1)**, suggesting that the use of simple descriptors in steps (1)-(6) was highly effective in winnowing the compound space to a tractable number of DFT calculations, while also retaining a considerable percentage of promising candidates. Next, *ab initio* molecular dynamics (AIMD) simulation was used to estimate $E_m$ (**Methods 2,** step (8)). The AIMD calculations are quite slow and as of this writing 26 out of the 80 compounds were studied and ranked by their estimated $E_m$ (**Table S2)**. 9 compounds with the estimated $E_m \leq 0.86$ eV were selected for more accurate determination of $E_m$ with long-time AIMD simulation. From these extended runs, 3 compounds with $E_m \leq 0.5$ eV were identified as members of promising new families of interstitial oxygen conductors, which were $K_2Mn_2(MoO_4)_3$, $La_4Mn_5Si_4O_{22}$, and $CeMn_2Ge_4O_{12}$. These three compounds are members of families of double molybdates $A_2TM_2(MoO_4)_3$ (A=alkali metal, TM=transition metal),[29] perrierite/chevkinite $RE_4TM_5Si_4O_{22}$ (RE=rare earth, TM=transition metal),[30] and germinates $RE_1TM_2Ge_4O_{12}$ (RE= rare earth, TM=transition metal),[31] respectively. These new families are a conservative estimate of the true number which could be uncovered with our present approach as many compounds have not yet been fully studied by AIMD and no effort was made to explore oxide structures not available in the Materials Project. $La_4Mn_5Si_4O_{22}$ (LMS) was selected as a representative member of the perrierite/chevkinite family for experimental investigation due to its predicted fast interstitial oxygen diffusion, simple established synthesis method, and inclusion of inexpensive, earth-abundant, and non-toxic elements, which are all traits desirable for new oxygen-active materials potentially useful in a variety of applications.

## Structure of $La_4Mn_5Si_4O_{22+\delta}$

LMS was first reported by Gueho *et al.* with structure and magnetic data in 1995,[32] with no further studies of which we are aware. LMS is a layered sorosilicate material with multivalent manganese and isostructural with perrierite and chevkinite, which crystallizes in the space group *C2/m*. In **Fig. 3a**, LMS displays eclipsed sorosilicate $Si_2O_7$ groups separated by rutile-like sheets of edge-shared $Mn_1^{4+}/Mn_2^{3+}$ octahedra and single isolated $Mn_3^{2+}$ octahedra. The stochiometric primitive cell has two $Mn_1^{4+}$, two $Mn_2^{3+}$, and one $Mn_3^{2+}$, where their valence states assignment fully consistent with the magnetic moments observed in the DFT calculations (**SI Discussion 3**). Sorosilicate $Si_2O_7$ groups show a zigzag arrangement along the a-axis, connect with $Mn_3^{2+}$ octahedra along b-axis and $Mn_2^{3+}$ octahedra along c-axis by sharing corners, leaving free space in between these unconnected $Si_2O_7$ chains. The La atoms are between the rutile-like layer and the sorosilicate layer, surrounded by 10 oxygen atoms. The structure has ample free space, a highly flexible network, and multiple Mn ions potentially capable of oxidation, making it ideally suited to form and transport interstitial oxygen.

## A dual diffusion mechanism enabled by undercoordinated sorosilicate groups and flexible corner-sharing framework

*Ab initio* studies and simple thermodynamic considerations suggest that excess oxygen with $La_4Mn_5Si_4O_{22+\delta}$ ( $\delta \approx 0.5$ ) is thermodynamically favorable under air conditions, while oxygen vacancies are very unfavorable due to their high formation energy (**SI Discussion 2**). In **Fig. 3a**, the most stable interstitial site ($O_i^1$) lies in between two adjacent sorosilicate $Si_2O_7$ groups, connecting two Si tetrahedra, and the second most stable interstitial site ($O_i^2$) lies in the joint of the $Si_2O_7$ and the $Mn_3^{2+}$ octahedra. These two prevailing interstitial sites contribute two distinct and competitive diffusion pathways observed from AIMD simulations. In the interstitial diffusion mechanism (yellow arrow in **Fig. 3a**), the $O_i$ hops between the $O_i^1$ sites through the channel between sorosilicate chains along the a-axis, with a barrier of 0.45 eV calculated by Climbing Image Nudged Elastic Band (CI-NEB) method (**Fig. S2a**). A parallel active interstitialcy (cooperative "knock-on") mechanism is indicated by cyan arrows, in which the $O_i$ moves along the corner-sharing $Si_2O_7$-$MnO_2$-$Si_2O_7$ framework along the b-axis. In the interstitialcy mechanism, the $O_i$ first hops from the $O_i^1$ site to a lattice site by kicking a lattice oxygen to the $O_i^2$ site, which then moves to a lattice site by kicking another lattice oxygen to the next $O_i^1$ site. By passing through the metastable interstitial $O_i^2$ site, the $O_i$ diffuses through an interstitialcy mechanism with a CI-NEB calculated barrier of 0.53 eV (**Fig. S2b**). The oxygen diffusion coefficient was calculated by AIMD simulations and machine learning-trained interatomic potential molecular dynamics (ML-IPMD) simulation (**Methods 3-4**), where the ML-IPMD was used to obtain more accurate diffusion coefficient through better sampling at lower temperatures compared to AIMD. **Fig. 3b** shows a calculated migration barrier of 0.44 eV over a wide temperature range, consistent with the CI-NEB barriers. The DFT predicted stable interstitials with low migration barriers indicate that $La_4Mn_5Si_4O_{22+\delta}$ is a fast oxygen conductor.

Nearly all known interstitial oxygen conductors diffuse oxygen ions through an interstitialcy mechanism. The prevalence of the interstitialcy mechanism is likely due to the large size of the oxygen anion and the lack of available diffusion channels. A pure interstitial diffusion mechanism through channels is rare, and, to our knowledge, only found in apatites. LMS adopts a dual-diffusion mechanism, where the oxygen diffusion is dominated by both interstitial and interstitialcy pathways. The pure interstitial diffusion channel between the sorosilicate chains has a slightly lower energy barrier than the interstitialcy pathway by a difference of 0.08 eV according to the CI-NEB calculations. The interstitial diffusion occurs due to the free space between the zigzag-arranged undercoordinated sorosilicate chains and the adjacent $Mn_3^{2+}$ ion, which create a stable channel for the interstitial oxygen. The interstitialcy diffusion happens through the rotation of $Si_2O_7$-$MnO_2$-$Si_2O_7$ polyhedral units, suggesting that the flexible corner-sharing framework may help stabilize the interstitial oxygen and facilitate diffusion. This dual mechanism is distinct from any confirmed mechanisms for an interstitial oxygen conductor of which we are aware.

## Experimental confirmation of interstitial oxygen conductor $La_4Mn_5Si_4O_{22+\delta}$

To validate the computational predictions, we synthesized LMS and studied its oxygen concentration, conductivity, and oxygen surface exchange rate. A molten salt synthesis method was used (**Methods 5**). The monoclinic perrierite (C2/m) structure of LMS was confirmed by Rietveld-refinement of room temperature X-ray diffraction (XRD) data (**Fig. 4a**), and the consistency of the obtained lattice parameters with previous literature values (**Table S4**). LMS pellets were sintered at 1050 °C for 24h for the elemental and electrical characterizations. The field emission scanning electron microscope (FESEM) image of sintered LMS pellet (**Fig. S4**) confirmed its high density with minimal porosity through the sample. The composition was determined as $La_{4.00}Mn_{4.69}Si_{4.03}O_{22.42}$ by electron probe micro-analyzer (EPMA) analysis, suggesting significant Mn deficiency, small Si excess, and significant excess oxygen with $\delta = 0.42$. Interstitial oxygen was also verified by the iodometric titration method with $\delta = 0.47$ (**Table S6, SI Discussion 5**). X-ray photoemission spectroscopy (XPS) analysis qualitatively supported the presence of interstitial oxygen (**Fig. 4b**). Deconvolution of the core level O1s spectra reveals three peaks. The peaks at 530.4 eV and 532.1 eV are associated with the lattice oxygen[33,34] and surface chemisorbed oxygen,[35] respectively. The low binding energy peak at 528.9 eV represents the interstitial oxygen within LMS.[33] The XPS survey scan on the LMS pellet surface revealed the presence of La, Mn, Si, and O without any detectable impurity elements (**Fig. S5**). Thermogravimetric analysis (TGA) was performed to study the temperature dependence of oxygen content (**Fig. 4c**). Detectable but modest reversible changes in oxygen content were observed when heating/cooling the LMS powder. The interstitial oxygen content $\delta$ changed from 0.42 (based on EPMA result) to 0.52, corresponding to a 0.5% change in the total oxygen content. The small change of oxygen content with respect to temperature is consistent with the *ab initio* results of $\delta \approx 0.5$, suggesting that the interstitials are stabilized by oxidizing the $Mn^{2+}$ ions and further oxidization of the system is difficult (**SI Discussion 2**). The TGA and *ab inito* results suggest LMS has a small thermodynamic factor, discussed more below.

LMS is a semiconductor with a narrow indirect band gap of 0.79 eV measured by UV-vis spectroscopy (**Fig. S6**), indicating that LMS could have electron conduction at high temperatures due to thermal excitation. The band gap was calculated by *ab initio* method with various functionals, and the predicted band gap of 0.72 eV using the strongly constrained and appropriately normed (SCAN) functional is the most consistent with experiment (**Table S3, SI Discussion 1**). The 4-probe conductivity measurement shows that LMS is a mixed ionic-electronic conductor (**Fig. S7**). Ionic conductivity ($\sigma_{ion}$) of LMS was measured using electron blocking 8YSZ,[36] and our setup was verified by obtaining robust results on the well-studied mixed conductor $La_{0.6}Sr_{0.4}Co_{0.2}Fe_{0.8}O_3$ (LSCF) (**Methods 9** and **Fig. S8-S9**). In **Fig. 4d,** LMS has comparable ionic conductivity to many of the best fast oxygen conductors and considerably higher ionic conductivity than widely used commercial materials such as LSCF and YSZ. In addition, LMS has comparable or improved ionic conductivity compared to other interstitial oxygen conductors, such as hexagonal perovskite $Ba_7Nb_{3.9}Mo_{1.1}O_{20.05}$, apatite $La_{9.75}Sr_{0.25}Si_6O_{26.895}$, Ruddlesden-Popper $Nd_2NiO_{4+\delta}$, and melilite $La_{1.54}Sr_{0.46}Ga_3O_{7.27}$. The experimental activation barrier of oxygen ion conduction in LMS is 0.72±0.03 eV, determined by the Arrhenius relation $\sigma T = \sigma_0 \mathrm{e}^{(-\frac{E_A}{k_b T})}$ (**Methods 11**). Given the small temperature dependence of oxygen stoichiometry from TGA, this experimental $E_A$ is expected to be similar to the DFT SCAN CI-NEB calculated migration barriers, which were 0.69 eV and 0.74 eV for interstitial and interstitialcy diffusion, respectively (**Fig. S2c, d**). Note that the above AIMD barrier is compared to GGA CI-NEB barriers as they both were simulated with GGA. Here the experimental barrier is compared to SCAN CI-NEB barriers, as the latter is expected to be the most accurate as we have calculated (**SI Discussion 1**). The good agreement between the experimental activation energy and DFT migration barriers further supports the dominance of both interstitial and interstitialcy mechanisms for oxygen diffusion in LMS.

To further probe the oxygen kinetics in LMS, the oxygen surface exchange coefficient ($k_{chem}$) and the chemical oxygen diffusivity ($D_{chem}$) were studied using the electrical conductivity relaxation (ECR) method. The Arrhenius plots of $D_{chem}$ and $k_{chem}$ of LMS in **Fig. 4e-f** show that LMS has $D_{chem}$ and $k_{chem}$ comparable to numerous state-of-art solid oxide electrode materials over a wide range of temperatures. LMS has among the highest $k_{chem}$ values at low temperatures of any known material, in part due to the relatively small activation energy of 0.82 eV for oxygen surface exchange (**Methods 11**). The enhanced $k_{chem}$ of LMS at low temperatures compared to vacancy-mediated diffusers might be due to there being more surface interstitial sites than vacancy sites available for oxygen exchange, but further study is needed to understand the interstitial surface exchange mechanism.

## Discussion

The oxygen tracer diffusion coefficient ($D^*$) was determined using the Nernst-Einstein equation (**Methods 11**). The thermodynamic factor $\gamma$ and the tracer surface exchange coefficient ($k^*$) were evaluated by $D_{chem} = \gamma D^*$ and $k_{chem} = \gamma k^*$ .[37] The thermodynamic factor of LMS varies from 21 to 28 in the temperature range of 600 to 750 °C (**Fig. S11f**), which is about 10 to 20 times smaller than that of the commonly studied mixed ionic electronic conducting perovskites LSCF,[38] BCFZr,[39] and BSCF.[40] The small $\gamma$ and high $k_{chem}$ imply LMS will have high $k^*$ compared with other state-of-the-art materials. **Fig. S13** shows that LMS has a higher $k^*$ at intermediate and low temperatures than state-of-the-art materials, perhaps even exceeding that of the leading BSCF material. The high surface exchange rate suggests LMS has the potential to assist in achieving fast oxygen reduction kinetics, e.g., reducing area specific resistance in the air electrode of solid oxide cells for electricity or hydrogen production, although its low electronic conductivity means it would have to be used as a composite with a good electrical conductor in many applications.[41]

This work demonstrates the largely untapped potential of interstitial oxygen ion conductors for reduced temperature oxygen-active material applications. We have designed a simple but effective method using physically-motivated descriptors and *ab initio* calculations to search for new families of interstitial oxygen diffusers. The effectiveness of our approach was confirmed by prediction and experimental confirmation of an entirely new class of predominantly interstitial oxygen conductor, represented by $La_4Mn_5Si_4O_{22+\delta}$ (LMS). LMS has very fast oxygen surface exchange and transport through both interstitial and interstitialcy mechanisms, which are enabled by the free space between sorosilicate chains, corner-sharing framework $Si_2O_7$-$MnO_2$-$Si_2O_7$, and high redox activity of the nearby $Mn_3^{2+}$ ion. We note that LMS is just one example of the broader perrierite/chevkinite $RE_4TM_5Si_4O_{22}$ structural family, other compositions within this family or composition and/or microstructure refinement of LMS itself may yield additional materials with fast oxygen interstitial transport. We also proposed that families of double molybdates $A_2TM_2(MoO_4)_3$ and germinates $RE_1TM_2Ge_4O_{12}$ are promising for further study. The success of the approach suggests that we have identified relatively simple structural and chemical features that strongly correlate with stable interstitial oxygen formation and fast migration. The discovery of a new high-performing interstitial oxygen conductor from experimental study of just one material suggests that the method is highly effective, and that many more materials exhibiting fast oxygen interstitials kinetics can be found through its application.

## Figures

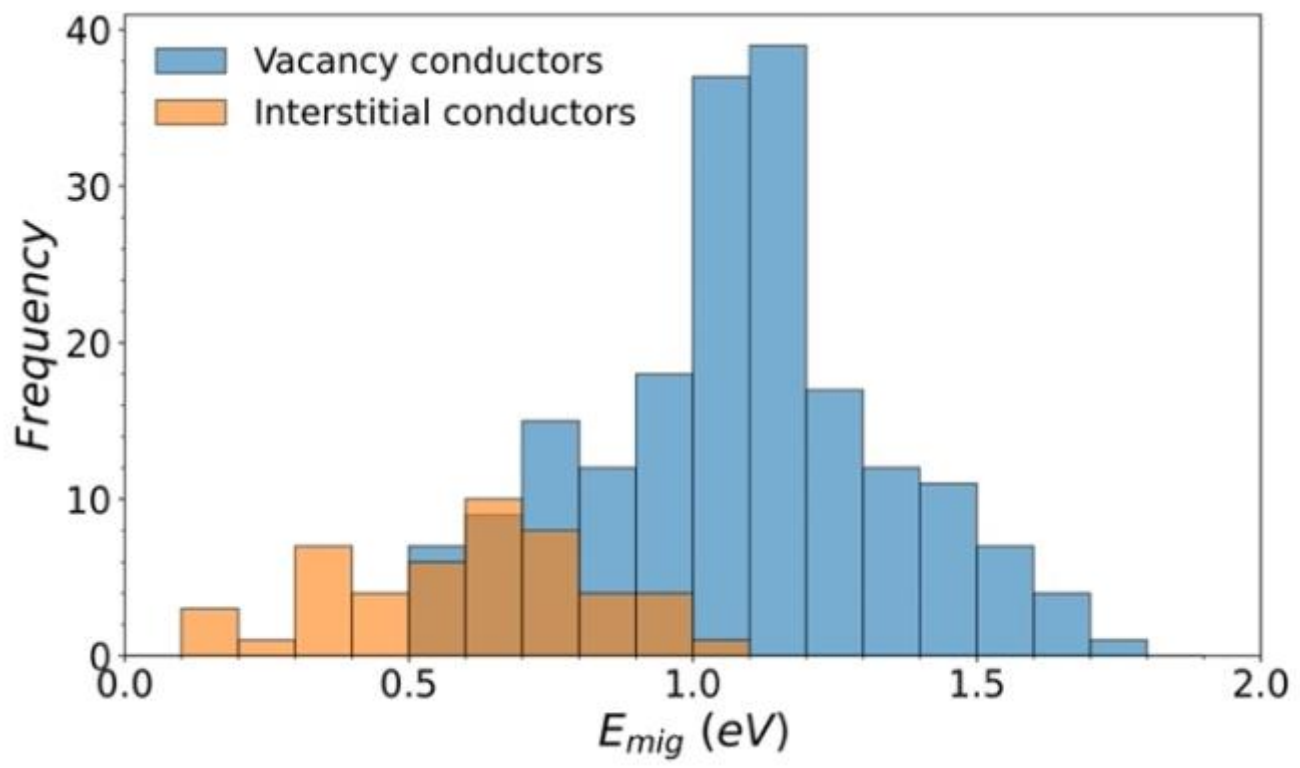

**Figure 1.** Migration barrier comparison of known interstitial and vacancy diffuser oxides obtained from the Citrine Informatics database[13] and literature. Data is available in digital Supplementary files.

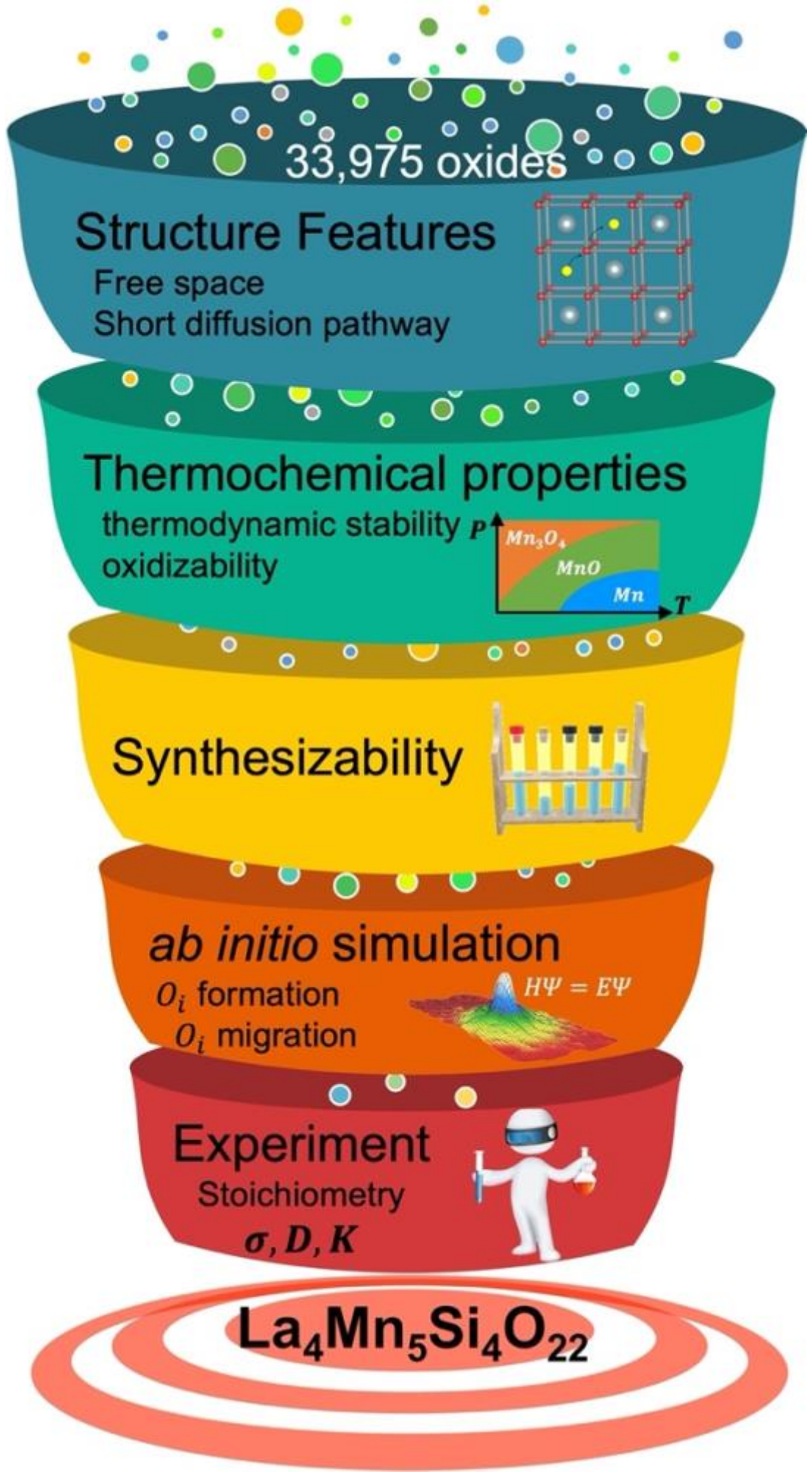

**Figure 2.** Schematic diagram of the screening approach for new interstitial oxygen conductors.

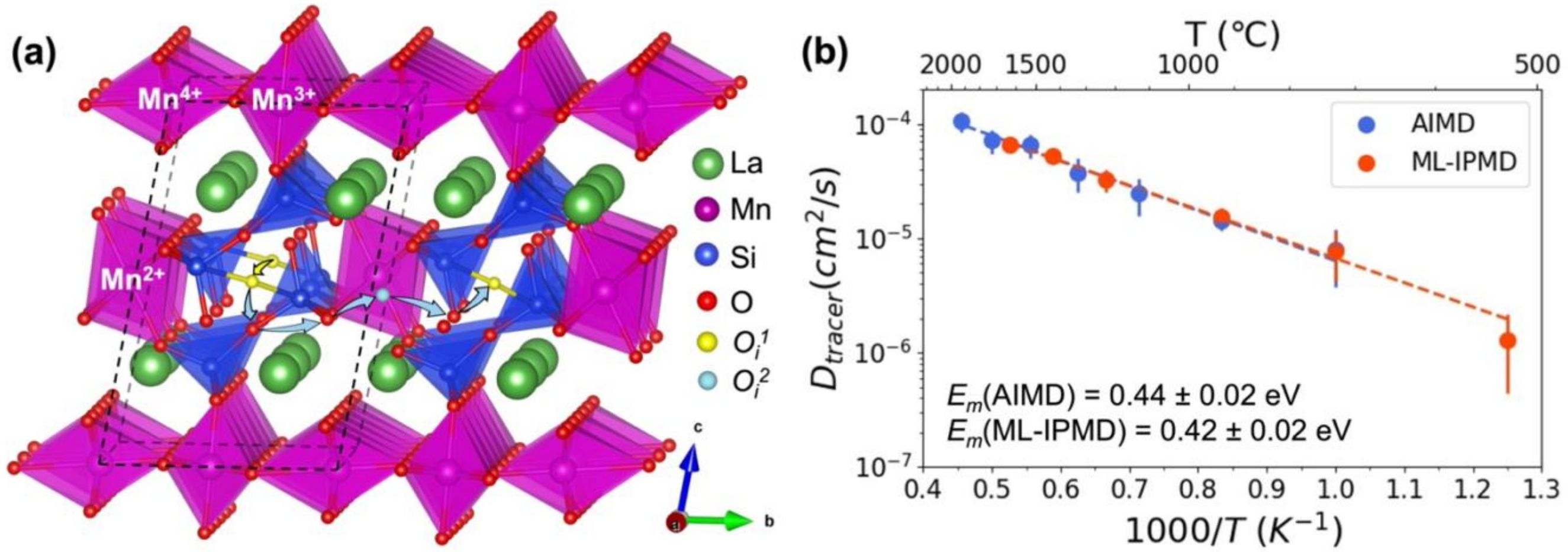

**Figure 3.** (a) Bulk structure of the $La_4Mn_5Si_4O_{22}$. The La, Mn, Si, and O sites are shown as green, purple, blue, and red spheres, respectively. The black dash line denotes the single unit cell. The yellow ball represents the $O_i^1$ site and the cyan ball represents the $O_i^2$ site, respectively. Interstitial Oxygen ($O_i$) in LMS diffuses through both interstitial mechanism (yellow arrow) and interstitialcy (cooperative "knock-on") mechanism (cyan arrows). (b) Arrhenius plot of tracer diffusivity $D_{tracer}$ of oxygen predicted by *ab initio* molecular dynamic (AIMD) and machine learning trained interatomic potential molecular dynamic (ML-IPMD) simulations. The error bars represent the standard deviation of $D_{tracer}$ determined by multi-origin analysis (**Methods 3**).

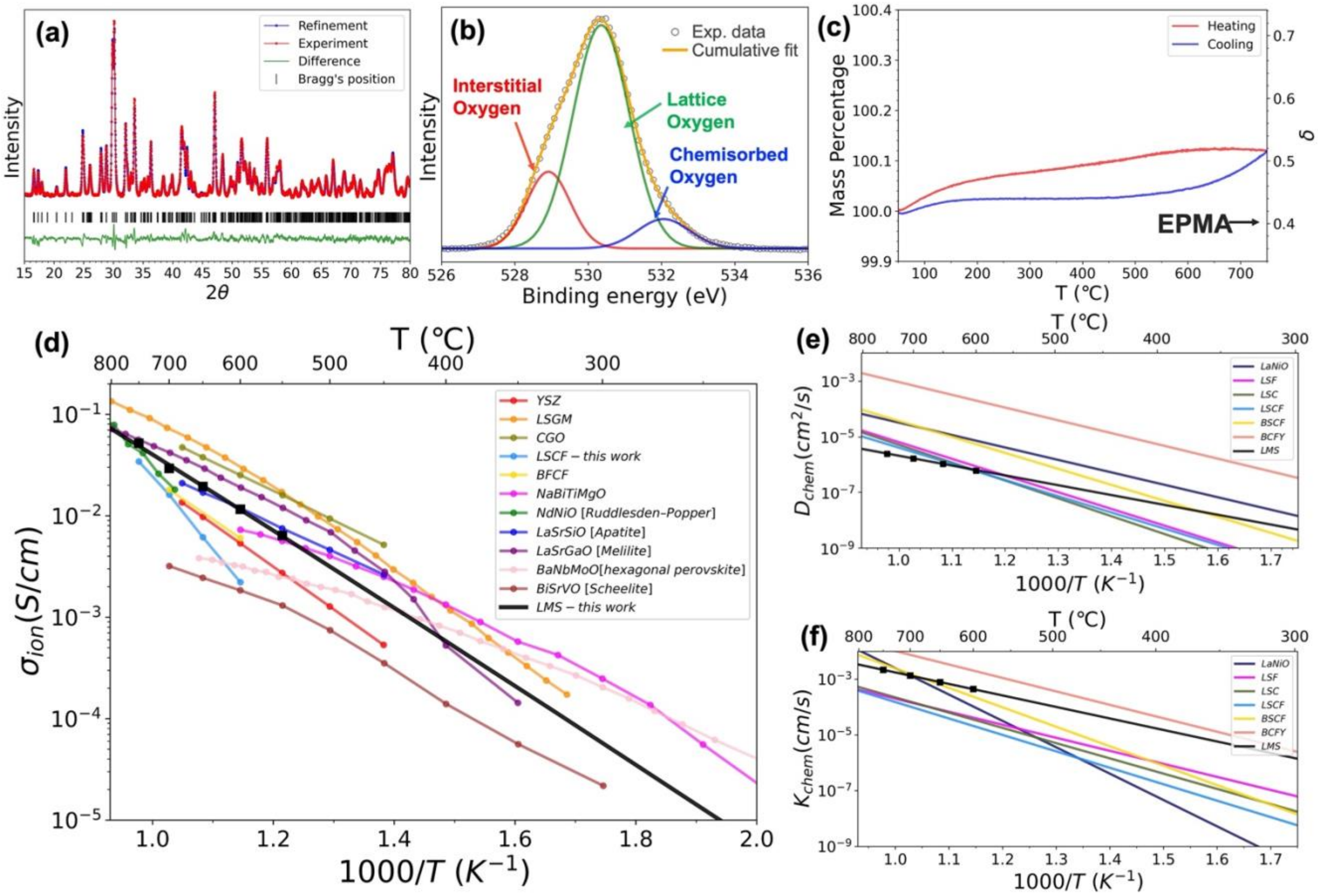

**Figure 4.** (a) Room temperature X-ray diffraction pattern of $La_4Mn_5Si_4O_{22+\delta}$ (LMS) and corresponding Rietveld refinement. (b) The peak-fitting results of O1s XPS spectra. (c) Mass percentage and corresponding interstitial oxygen content ($\delta$) change during the thermogravimetric analysis of LMS between 50 °C and 750 °C under a 1 atm $O_2$ environment, inserted arrow points the interstitial oxygen contents of $\delta$=0.42 measured by the electron probe micro-analyzer (EPMA) analysis. (d) Arrhenius plots of the measured ionic conductivity of LMS compared with the leading oxygen ion conductors $Zr_{0.92}Y_{0.08}O_{2-\delta}$ [YSZ],[20] $La_{0.8}Sr_{0.2}Ga_{0.83}Mg_{0.17}O_{3-\delta}$ [LSGM],[42] $Ce_{0.9}Gd_{0.1}O_{1.95}$ [CGO],[43] $(La_{0.6}Sr_{0.4})_{0.95}Co_{0.2}Fe_{0.8}O_{3-\delta}$ [LSCF - this work], $Ba_{0.5}Sr_{0.5}Co_{0.8}Fe_{0.2}O_{3-\delta}$ [BSCF],[44] $Na_{0.5}Bi_{0.49}Ti_{0.98}Mg_{0.02}O_{2.965}$ [Mg-doped NBT perovskite],[45] $Nd_2NiO_{4+\delta}$ [Ruddlesden-Popper],[23] $La_{9.75}Sr_{0.25}Si_6O_{26.895}$ [Apatite],[16] $La_{1.54}Sr_{0.46}Ga_3O_{7.27}$ [Melilite],[24] $Ba_7Nb_{3.9}Mo_{1.1}O_{20.05}$ [hexagonal perovskite],[25] and $Bi_{0.975}Sr_{0.025}VO_{3.9875}$ [Scheelite].[46] (e) $D_{chem}$ and (f) the $k_{chem}$ of LMS comparing with $La_2NiO_{4+\delta}$ (LaNiO), $La_{0.5}Sr_{0.5}FeO_3$ (LSF), $La_{0.5}Sr_{0.5}CoO_3$ (LSC), $La_{0.6}Sr_{0.4}Co_{0.2}Fe_{0.8}O_3$ (LSCF), $Ba_{0.5}Sr_{0.5}Co_{0.8}Fe_{0.2}O_3$ (BSCF), and $BaCo_{0.7}Fe_{0.22}Y_{0.08}O_{3-\delta}$ (BCFY).

1

## Methods

### 1. Descriptor screening approach

Step (1) Free space. Crystallographic information files (CIF files) from the Materials Project database were used to calculate the free volume for forming interstitials. DFT-computed structures in the Materials Project are first analyzed using the Voronoi analysis method implemented in pymatgen[1] to find the potential interstitial sites. Distance from the interstitial site to its nearest neighboring cations $d_c$ and anions $d_a$ were used as the descriptor of the free volume needed for an interstitial oxygen anion $O^{2-}$. Specifically, the screening criteria were established as $d_c \geq 0.99$ Å and $d_a \geq 0.88$ Å. The distance criteria applied above were established by taking the minimum distances observed for the interstitial oxygens in Ruddlesden-Popper $La_2NiO_4$ and apatite $La_{10}Si_6O_{27}$, which were taken as a guide to determine if there is sufficient room for an interstitial oxygen anion $O^{2-}$ in the material. Interstitial sites that met the screening criteria were identified as reasonable interstitial oxygen sites $O_i$ for the next step and only structures with such $O_i$ sites were retained. After step (1), 16,455 compounds were retained, and the search space was reduced by 52%.

Step (2) Short hop distance. The minimum distance between the two nearest neighboring $O_i$ sites from step (1) was used as the criterion representing the hop distance. Only materials with this distance ≤ 3 Å were retained. This distance was chosen based on the shortest hop distance of interstitial oxygen in Ruddlesden-Popper $La_2NiO_4$ and apatite $La_{10}Si_6O_{27}$, for which the hop distance is ~2.8 Å and ~2.7 Å, respectively. After step (2), 9,477 compounds were retained, and the search space was reduced by 72%.

Step (3) thermodynamic stability. Materials stability was assessed based on energy relative to the convex hull ($E_{hull}$) as computed by the Materials Project. Only compounds with $E_{hull}$ < 100 meV/atom (closed system) and $E_{hull}$ < 200 meV/atom (system open to $P(O_2)$ = 0.2 atm and T = 300 K) were retained. These thresholds were chosen to be simple round numbers and within a few tens of meV/atom above the cutoffs that identify the Ruddlesden-Popper $La_2NiO_4$ and apatite $La_{10}Si_6O_{27}$ as acceptably stable, since these are well-studied interstitial oxygen conductors known to be reasonably stable under many device conditions. Ruddlesden-Popper $La_2NiO_4$ has $E_{hull}$ = 76 meV (closed system) and $E_{hull}$ = 94 meV (open to $P(O_2)$ = 0.2 atm and T = 300 K), and apatite $La_{10}Si_6O_{27}$ has $E_{hull}$ = 54 meV (closed system) and $E_{hull}$ = 169 meV (open to $P(O_2)$ = 0.2 atm and T = 300 K), respectively.

Step (4) oxidizability. The valence state of each site was predicted by the Bond Valence analysis[2] module incorporated in pymatgen. To ensure the redox process of including additional oxygen into the materials with a low barrier, we proposed that the redox property from the cations is critical. If the valence state of a cation is smaller than its maximum oxidation state, for example, the Mn atom is at 2+ instead 7+, there are free valence electrons associated with this cation, and the compound has high redox potential and thus can be easily oxidized. Oxidizability is assessed by calculating the sum of the differences between the highest valence state and the actual valence state of each cation ion. Only compounds predicted to have oxidizable cations were retained.

Step (5) synthesizability. Materials are checked for whether it was reported as being synthesized at least once in the Inorganic Crystal Structure Database (ICSD)[3]. Compounds were retained only if the structure has an entry of experimental data in ICSD. After steps (3)-(5), only 519 compounds were retained, and the search space was reduced by 98.5%.

Step (6) Structure similarity analysis. For the remaining 519 compounds, a structure similarity analysis[4] assessed based on local coordination information from all sites in two structures was performed to group these compounds into 345 structural families. One candidate for each structure group was selected for further validation with *ab initio* studies. The selection of the candidate material followed a two-step process. First, the candidate was selected from the top of the stability ranking. If multiple materials shared the same level of stability, the candidate was subsequently selected by its oxidizability ranking.

### 2. *Ab initio* simulation screening approach

Step (7) Formation energy ($E_f$). Density functional theory (DFT) calculations were used to calculate the formation energy ($E_f$). The interstitial site with the largest free space is calculated. Structure optimization was performed for the bulk structures from the Materials Project. Supercells were built from the relaxed bulk structure with an approximately minimum length of 8 Å along each direction to minimize defect self-interaction. Formation energy was calculated by

$E_f = E_{(supercell\ with\ Oi)} - E_{(supercell)} - \mu_O$ (1),

where the $\mu_O$ is the oxygen chemical potential under air condition. The O chemical potential was calculated by using a combination of DFT calculated total energies and experimental thermodynamic data for $O_2$ gas at air condition[5] with the following form[6]

$$\mu_O = \frac{1}{2}\left[E_{O_2}^{VASP} + \Delta h_{O_2}^0 + H(T, P^0) - H(T^0, P^0) - TS(T, P^0) + kT\ln\left(\frac{P}{P^0}\right) - \left(G_{O_2}^{s,vib}(T) - H_{O_2}^{s,vib}(T^0)\right)\right] \quad (2),$$

where $E_{O_2}^{VASP}$ is the *ab initio* calculated energy of an $O_2$ gas molecule, $\Delta h_{O_2}^0$ is a numerical correction that takes into account the temperature increase of $O_2$ gas from 0 K to $T^0$, the contribution to the enthalpy at $T^0$ when oxygen is in the solid phase, and the numerical error in overbinding of the $O_2$ molecule in DFT. $\Delta h_{O_2}^0$ is obtained from comparing calculated formation energies and experimental formation enthalpies of numerous oxides, we used $\Delta h_{O_2}^0$ =0.70 eV/O from Ref. 7. $H(T, P^0)$ and $H(T^0, P^0)$ are the gas enthalpy values at standard and general temperatures $T^0$ and $T$, respectively. In this case, $T^0$ is 298 K and $T$ refers to the room temperature 300K. $TS(T, P^0)$ is the gas entropy, and the logarithmic term is the adjustment of the chemical potential for arbitrary pressure, where $P$ and $P^0$ are the referenced pressure and the standard pressure, respectively. In this case, the referenced pressure is 0.2 atm. The $\left(G_{O_2}^{s,vib}(T) - H_{O_2}^{s,vib}(T^0)\right)$ term accounts for the solid phase vibrations, which are approximated with an Einstein model with an Einstein temperature of 500 K.[8]

Step (8) Migration barrier $E_m$. *Ab initio* molecular dynamics (AIMD) was performed at 2000K for 30ps for the remaining 80 compounds to have an initial evaluation of the migration barrier $E_m$. We used this initial run to observe the AIMD trajectories for hopping times and estimate the migration barrier based on the hop rate $r$

$$r = v\mathrm{e}^{(-\frac{E_m}{k_b T})} \quad (3),$$

where $v$ is the attempt frequency, estimated as $5\times10^{12}$ $s^{-1}$, $k_b$ is the Boltzmann constant, and $T$ is the temperature, respectively. The hop rate $r$ was observed from the AIMD simulation, and an initial estimation for $E_m$ could be determined. The hop rate of 0.033/ps, i.e., 1 hop observed within 30ps AIMD simulation, corresponds to a migration barrier of 0.86 eV. Materials that have at least 1 hop observed within 30ps at 2000K, were selected for further DFT studies with long AIMD runs at different temperatures to calculate oxygen tracer diffusion coefficients, which were fit to an Arrhenius form $Ae^{E_m/kT}$.

All the calculations in this screening approach were performed with DFT using the Vienna ab Initio Simulation Package (VASP) code.[9] The generalized gradient approximation exchange-correlation functional Perdew, Burke, and Ernzerhof (GGA-PBE)[10] and projector augmented wave method (PAW)[11] were used for the effective potential for all atoms. The valence electron configuration of the La, Mn, Si, and O atoms utilized in all calculations were $5s^25p^66s^24d^1$, $3p^64s^23d^5$, $3s^23p^2$, and $2s^22p^4$, respectively. The plane wave cutoff energy was 520 eV and spin-polarized calculations were performed. For the defect formation energy calculations, the stopping criteria for total energy calculations were 0.01 meV/cell for the electronic relaxation and 0.05 eV/Å for ionic relaxation, respectively. K-point meshes were automatically generated based on the structural volume with a k points density of 0.04/Å$^{-3}$ to ensure calculation accuracy. For the AIMD simulations were performed using gamma-point-only sampling of k-space. The structure was first heated up to 2000K within 0.3 ps in the NVT ensemble using the Andersen thermostat, and then simulated in the NVT ensemble using a Nosé–Hoover thermostat[12,13] for 30ps.

### 3. AIMD simulation of $O_i$ diffusion in $La_4Mn_5Si_4O_{22+\delta}$

With approximately 2% $O_i$ concentration, the oxygen ion diffusivity in $La_4Mn_5Si_4O_{22+\delta}$ was studied by *ab initio* molecular dynamic simulation at the temperature from 1000K to 2200K, with a step of 200K. The structure was first heated up to 2000K within 0.3 ps in the NVT ensemble using the Andersen thermostat, and then simulated in the NVT ensemble using a Nosé–Hoover thermostat[12,13] for 300ps at each temperature state. No signs of melting at high temperatures were observed. No effort was made to correct for thermal expansion as it was assumed the effect would be small. We performed a multi-time origin method[14] to calculate the average diffusion coefficient $D$ along with its standard deviation to ensure a good statistical averaging. The diffusion coefficient within a simulation time $t$ was evaluated using the mean squared displacement by

$$D_t = \left[\frac{1}{6t}\langle MSD\rangle\right]_t \quad (4),$$

where $t$ is the simulation time. We evaluate $D_{t_i}$ for all times between $t_i$ to $t_i$+120ps, where $t_i$ changes from 0 to 180ps with a step of 1.2ps. Then, the diffusion coefficient $D$ and its standard deviation $D_{std}$ were calculated by all the sampled $D_{t_i}$ by $D = \frac{\sum D_{t_i}}{n}$ and $D_{std} = \sqrt{\frac{\sum(D-D_{t_i})^2}{n-1}}$, where $n$ is the total number of $t_i$.

Then, the migration barrier was calculated by fitting the Arrhenius relationship using the diffusion coefficient at the 5 temperatures by

$$D = D_0 \mathrm{e}^{(-\frac{E_m}{k_b T})} \quad (5),$$

where $D_0$ is the pre-exponential factor, $k_b$ is the Boltzmann constant, and $T$ is the temperature. Finally, the ionic conductivity was calculated from the diffusion coefficient based on the Einstein-Nernst equation,

$$\sigma = \frac{cz^2F^2}{RT} D \quad (6),$$

where $c$ is the volume concentration of oxygen species, $z$ is the charge of each oxygen ion, $F$ is the Faraday constant, $R$ is the gas constant, and T is temperature.

**4. ML-IPMD simulations.**

Machine learning interatomic potential (ML-IP) was trained by the moment tensor potential (MTP) method[15,16]. The training data were obtained from the AIMD trajectories from 1000K to 2600K with an interval of 200K. For each temperature, we collected 200 structures with a time interval of 0.12ps to cast aside similar structures. An optimized MTP is then obtained by minimizing the errors in the predicted energies, forces, and stresses with respect to the DFT data. We set the weights of 100:1:0 to the energy, force, and stress data points, following previous works.[17,18] The radius cutoff was set to be 5.0 Å, a typical value used in previously reported MTPs, and the maximum level of the basis functions is set to be 20. All the training and evaluations were performed using the Machine-Learning Interatomic Potentials (MLIP) package.[19] Classical MD simulations were performed using the trained MTP. The time step was set to 1 fs, and the total simulation time was 1 ns for temperatures above 1200K, 5 ns at 1000K, 20ns at 800K, and 100ns at 600K, respectively.

**5. Synthesis.**

Stoichiometric quantities of highly pure $La_2O_3$ (Alfa Aesar, 99.99 %), $MnO_2$ (Acros Organics, > 99.99 %), and $SiO_2$ (Alfa Aesar, 99.9 %) were mixed with KCl (Alfa Aesar, 99 – 100.5 %) flux. The flux-to-reactants molar ratio was 28.5:1 (mass ratio was 1.6:1). The reaction mixture was heated in a covered alumina crucible (30ml) at 900 °C for 6 days. The sample was slowly cooled down to 500 °C at a cooling rate of 20 °C/h and then further cooled down to room temperature at a cooling rate of 95 °C/h. The resultant mixture was washed with deionized water several times to remove the KCl flux, and then the obtained powder was dried on a hot plate at 120 °C in air. The LMS powder was grounded to fine powder in acetone medium using a porcelain mortar-pestle and pelletized in a rectangular bar using polyvinyl alcohol as binder and sintered at 1050 °C for 24 h.

**6. Characterization.**

The structural characterization of the LMS powder and pellet sample was performed using room temperature X-ray diffraction (XRD) technique using Cu-Kα source (Bruker D8 Discovery) followed by Rietveld refinement using Fullprof code.[20] The Field effect scanning electron microscopy (FESEM) image of the pellet was collected using a high-resolution microscope (Zeiss 1530). The UV-vis diffuse reflectance of the LMS pellet was measured using a spectrophotometer (Perkin Elmer Lambda 19 UV/Vis/NIR) and the optical band gap was determined using the Kubelka Munk equation[21] and Tauc plot.[22] X-ray photoemission spectroscopy analysis was performed on the sintered LMS pellet surface using a Thermo K-Alpha X-ray photoelectron spectrometer (Al-ka source). The pellet surface was cleaned by Ar sputtering inside the XPS chamber at ultra-high vacuum (~ $10^{-9}$ torr). The spot size of the X-ray beam was 400 micrometers.

**7. EPMA Analysis.**

The chemical composition of the polished LMS pellet was measured using a CAMECA SX-Five FE-EPMA operated at 12 kV accelerating voltage and 20 nA beam current. The lowest possible accelerating voltage was selected that would minimize activation volume while also providing adequate overvoltage for excitation of Mn K$\alpha$. Samples were mounted in epoxy and polished with colloidal alumina suspension. Samples and standards were coated with 1 nm Ir immediately prior to analysis. The O K$\alpha$ X-rays were measured with PC0 crystal (2d=47.12Å), Si K$\alpha$ with LTAP (thallium acid phthalate, large format), Mn K$\alpha$ with LLIF (lithium fluoride, large format), and La L$\alpha$ with LPET (pentaerythritol, large format). Pulse height analysis was operated in differential mode for O and Si to avoid high order reflections from La. Spectral resolution of the PC0 crystal on 160 mm radius Rowland circle was adequate to resolve O Ka peaks from interference with Mn L*l* without requiring direct interference correction. Measurements included a 20s peak counting time and 10s counting time for each high and low background position. The electron beam was fully focused for spot measurements, with a practical spot diameter of approximately 250 nm. The accuracy of individual measurements was evaluated based on the quality of the analytical total. The average atomic percentage was determined from the analysis of 14 individual grains on the sample surface with <0.5 wt% standard deviation for each element and used to calculate the bulk atomic formula (**Table S6**).

8. **Thermogravimetry analysis.**

Thermogravimetric analysis (TGA) was performed in oxygen at the temperature range from 50 to 750 °C at a heating and cooling rate of 3 °C/ min using a TGA analyzer (TA Instruments Q500). The ground LMS powder was heated in two cycles to remove any adsorbed species. The third cycle is plotted in **Fig. 4c**. The room temperature oxygen content for LMS was considered from the EPMA results. A small mass correction (% of total mass) was performed to compensate for the sudden mass jump at the beginning of the heating cycle. We speculate that this abrupt fluctuation is associated with the instrument heater as similar behavior was observed in other materials too. The structural stability of the sample was also confirmed by performing XRD analysis after the TGA measurements.

9. **Conductivity measurement.**

The total conductivity of the LMS was measured in air conditions by the conventional DC 4-probe method using a Keithley 6221 power supply and a 2182A nanovoltmeter. Pt wire and silver paste were used for the electrical connection. Pt wires were used as the current and voltage leads. The current leads were connected to the two ends of LMS pellet using Ag paste. The voltage leads were wrapped around the LMS pellet at equal distances from the two ends using Ag-paste filling the gap between the Pt wire and LMS pellet to ensure a good electrical connection. For all the conductivity measurements, we used the current in the range of 1 to 10 μA.

Electronic conductivity was measured using Au as the ionic blocking electrode at the two ends of the polished LMS pellet.[23] We first deposited Au (300 nm) on one cross-section of the LMS pellet using Au sputtering unit (Leica ACE600) and heated it at 600 °C for 2 hours, and then repeated for the other cross-section. Afterwards, we deposited a thick Au layer (approximately 200 microns) on both ends using high pure Au paste and heated it at 900 °C for 2 hours. Pseudo 4-probe method was used to measure the electronic conductivity, where both the current and voltage leads were connected to the Au electrode terminals.

The ionic conductivity measurement was performed using 8YSZ blocks as the electronic blocking electrode.[23] A schematic of the assembly was displayed in **Fig. S8** along with the details of the setup in the supplementary file. The ionic conductivity of LMS was measured in air from 600 to 750 °C. To measure the resistance at a particular temperature, we measured the voltage across the LMS pellet at different currents and determined the resistance from the linear region of the I-V curve as shown in **Fig. S10**. The non-linearity at the higher current region of the I-V curve indicates the presence of ionic conductivity in LMS.[23] Conductivity measurements were conducted on three LMS pellets, The averaged conductivity along with its standard deviation are displayed in **Fig. S7**.

10. **Electrical conductivity relaxation (ECR).**

ECR study was performed using LMS pellet of length ~15 mm, width ~6 mm and thickness ~0.6 mm. A vertical and sealed alumina tube was used as the ECR chamber. The sintered LMS pellet was kept floating inside the chamber using 4 Pt-wires passing through a 4-bore alumina tube. The excess space of the ECR chamber was filled using alumina balls of average diameter of ~ 3 mm to reduce the effective chamber volume for faster gas exchange. The same electrical connection method described in the total conductivity measurement was used. 21 % and 5 % oxygen balanced $N_2$ gas with a total flow of 300 SCCM was used to create different $P(O_2)$ conditions inside the chamber. A 4-way gas switching valve was installed for the fast gas switching at the sample chamber. Abrupt oxygen partial pressure was changed from 5 to 21 % and vice versa at 750, 700, 650 and 600 °C, and the transient conductivity was measured. The normalized and fitted ECR data during the oxidation and reduction at different temperatures were shown in **Fig. S11a-b**. The value of $D_{chem}$ and $k_{chem}$ of LMS at a fixed temperature was determined as the average during oxidation and reduction at that temperature. The fitting details were presented in **SI Discussion 7**.

**11. Fittings of experimental data**

The experimental activation energy of oxygen ion conducting in LMS was fitted on three separate measurements conducted on three LMS pellets using the Arrhenius relation $\sigma T = \sigma_0 \mathrm{e}^{(-\frac{E_A}{k_b T})}$, from which the average activation is 0.72 eV with a standard deviation of 0.03 eV from the three LMS pellets.

The experimental activation energies of chemical oxygen diffusivity $D_{chem}$ and surface exchange coefficient were obtained by fitting the Arrhenius relation $D_{chem} = D_{chem}^0 \mathrm{e}^{(-\frac{E_A}{k_b T})}$ and $k_{chem} = k_{chem}^0 \mathrm{e}^{(-\frac{E_A}{k_b T})}$. The $D_{chem}$ and $k_{chem}$ were obtained from ECR analysis on one LMS pellet. The activation energy is 0.70±0.01 eV for $D_{chem}$ and 0.82±0.01 eV for $k_{chem}$, respectively, where the error bars are the standard deviation, representing the goodness of fitting.

The tracer diffusion coefficient $D^*$ was derived by the Nernst-Einstein equation $D^* = \frac{\sigma RT}{cz^2F^2}$ using the experimental conductivity $\sigma$ and experimental oxygen volume concentration $c$ from TGA results.

## Data availability

Source data and data that support the plots within this paper are available on Figshare.

## Acknowledgements

This work was funded by the US Department of Energy (DOE), Office of Science, Basic Energy Sciences (BES), under Award # DE-SC0020419. This work used the Extreme Science and Engineering Discovery Environment (XSEDE), which is supported by National Science Foundation Grant Number ACI-1548562.

## Author contributions

J.M. and S.S. have equal contributions to this work. R.J, and D.M. conceived and managed the project. J.M. performed the screening, ab initio calculations, and theoretical analyses with assistance from D.M. and R.J. S.S performed the synthesis, characterization, conductivity, and kinetic measurements. J.L. helped with the ECR analysis and contributed to scientific discussions. W.O.N. performed the EPMA analysis. X.L. trained the machine learning interatomic potential. J.M. wrote the first version of the manuscript with input from S.S. R.J. and D.M. reviewed the manuscript. All authors have reviewed and commented on the manuscript.

## Competing interests

The authors declare no competing interests.

# Computational Discovery of Fast Interstitial Oxygen Conductors

Jun Meng[1,4], Md Sariful Sheikh[1,4], Ryan Jacobs[1], Jian Liu[2], William O. Nachlas[3], Xiangguo Li[1], Dane Morgan[1,*]

*[1] Department of Materials Science and Engineering, University of Wisconsin Madison, Madison, WI, USA.*
*[2] DOE National Energy Technology Laboratory, Morgantown, WV, USA.*
*[3] Department of Geoscience, University of Wisconsin Madison, Madison, WI, USA.*
*[4] These authors contributed equally: Jun Meng, Md Sariful Sheikh. E-mails: ddmorgan@wisc.edu.*

## Table of Contents

## Supplementary Tables

**Table S1.** Materials list with interstitial oxygen formation energy $E_f \leq 0.3$ eV under atmosphere condition (T=300 K, P($O_2$)=0.2 atm) by *ab initio* calculation.

| Materials-ID | Formula | $E_f$ | Materials-ID | Formula | $E_f$ |
|---|---|---|---|---|---|
| mp-23349 | BiB3O6 | -3.85 | mp-772957 | SrV4O10 | -0.42 |
| mp-1196071 | Ba2Fe2O5 | -3.51 | mp-1196110 | SrCuTe2O7 | -0.34 |
| mp-1204837 | NaFe2Si6O15 | -3.25 | mp-1200170 | Ba5Cr3O13 | -0.34 |
| mp-23356 | Bi4B2O9 | -2.73 | mp-23446 | GeBi2O5 | -0.33 |
| mp-555752 | NaFe2Mo3O12 | -2.68 | mp-1195799 | K2Fe2B2O7 | -0.32 |
| mp-1199587 | Yb2VO5 | -2.28 | mp-556076 | Sr2Co2O5 | -0.30 |
| mp-555924 | Ca5Nb5O17 | -2.10 | mp-29189 | VHg2O4 | -0.29 |
| mp-542931 | Bi2B8O15 | -2.05 | mp-667343 | Re2Hg5O10 | -0.29 |
| mp-559364 | SrBi2B4O10 | -2.00 | mp-1204772 | Co2As2O7 | -0.27 |
| mp-29508 | LiMo3O9 | -1.91 | mp-558472 | SrCu2B2O6 | -0.23 |
| mp-744682 | Cr8Bi4O29 | -1.69 | mp-705159 | K2Co2Mo3O12 | -0.23 |
| mp-29058 | V3Bi6O16 | -1.61 | mp-563010 | RbFeMo2O8 | -0.20 |
| mp-18907 | Ca2MnAlO5 | -1.55 | mp-6496 | Ba2NaCu3O6 | -0.17 |
| mp-1194512 | V2Cu3O9 | -1.47 | mp-29112 | CrHg5O6 | -0.15 |
| mp-630403 | Ca2MnGaO5 | -1.45 | mp-18924 | Sr3Fe2O6 | -0.15 |
| mp-558429 | NaFe4Mo5O20 | -1.37 | mp-558751 | CaBi2O4 | -0.14 |
| mp-19290 | Mn2As2O7 | -1.33 | mp-6027 | Ba2Tl2CuO6 | -0.08 |
| mp-559180 | Ba2CuB2O6 | -1.28 | mp-704097 | Sr6Co4Bi2O15 | -0.05 |
| mp-22113 | Ca2Fe2O5 | -1.25 | mp-1203275 | Co2AsO5 | -0.04 |
| mp-1003437 | KMn2O4 | -1.24 | mp-1203433 | Hg3AsO5 | -0.04 |
| mp-21926 | SrFe2O4 | -1.23 | mp-647862 | Cr2Mo3O12 | -0.04 |
| mp-753258 | Li3CrO4 | -1.13 | mp-29048 | SrBi2O4 | 0.00 |
| mp-19165 | BaFeSi4O10 | -1.06 | mp-1190373 | K2V2CoO7 | 0.01 |
| mp-29259 | Bi2PdO4 | -1.06 | mp-554698 | K10MnMo7O27 | 0.07 |
| mp-556203 | La8Ni4O17 | -1.06 | mp-17387 | LiVAsO5 | 0.09 |
| mp-20161 | Na2CoGeO4 | -0.91 | mp-18893 | Ca2Mn3O8 | 0.12 |
| mp-18096 | Na2CoSi4O10 | -0.91 | mp-559685 | V2Cd4Te3O15 | 0.15 |
| mp-18926 | La3Ni2O7 | -0.90 | mp-1200219 | V4Cr2O13 | 0.16 |
| mp-21635 | CeMn2Ge4O12 | -0.87 | mp-546111 | Cr3AgO8 | 0.18 |
| mp-1194618 | Ba4Y2Fe2O11 | -0.87 | mp-560340 | La2Pd2O5 | 0.18 |
| mp-19228 | K2MnV4O12 | -0.82 | mp-1200054 | V4Fe2O13 | 0.19 |
| mp-505042 | CuBi2O4 | -0.78 | mp-541433 | CdBi2O4 | 0.21 |
| mp-550998 | TiZnBi2O6 | -0.73 | mp-19142 | Mn2V2O7 | 0.23 |
| mp-541464 | La4Mn5Si4O22 | -0.70 | mp-639811 | KIrO3 | 0.24 |
| mp-558316 | La4Ni3O10 | -0.69 | mp-19395 | MnO2 | 0.25 |
| mp-37961 | MgV3O8 | -0.59 | mp-1105484 | Sm3FeO6 | 0.26 |
| mp-18456 | LaCrO4 | -0.59 | mp-560273 | Cu3WO6 | 0.28 |
| mp-557927 | Na2SrV3O9 | -0.56 | mp-777667 | LiV2O5 | 0.28 |
| mp-1191696 | LiCrMo2O8 | -0.56 | mp-553887 | NaV2Bi3O10 | 0.28 |
| mp-2669 | Mo8O23 | -0.52 | mp-1194927 | VCdCoO5 | 0.30 |
| mp-22427 | BaSrFe4O8 | -0.51 | | | |

**Table S2.** Materials list studied by initial evaluation by 30ps *ab initio* molecular dynamic simulation at 2000K.

| Materials-ID | Formula | *Estimated $E_m$* | Materials-ID | Formula | *Estimated $E_m$* |
|---|---|---|---|---|---|
| mp-651434 | K2Mn2Mo3O12 | 0.36 | mp-541433 | CdBi2O4 | structure not stable |
| mp-541464 | La4Mn5Si4O22 | 0.44 | mp-1003437 | KMn2O4 | structure not stable |
| mp-21635 | CeMn2Ge4O12 | 0.6 | mp-1200219 | V4Cr2O13 | structure not stable |
| mp-29259 | Bi2PdO4 | 0.63 | mp-19290 | Mn2As2O7 | structure not stable |
| mp-556076 | Sr2Co2O5 | 0.62 | mp-777667 | LiV2O5 | structure not stable |
| mp-19165 | BaFeSi4O10 | 0.74 | mp-753258 | Li3CrO4 | structure not stable |
| mp-18893 | Ca2Mn3O8 | 0.8 | mp-29112 | CrHg5O6 | structure not stable |
| mp-630403 | Ca2MnGaO5 | 0.83 | mp-1196110 | SrCuTe2O7 | structure not stable |
| mp-639811 | KIrO3 | 0.86 | mp-37961 | MgV3O8 | structure not stable |
| mp-647862 | Cr2Mo3O12 | >0.86 | mp-19395 | MnO2 | structure not stable |
| mp-17387 | LiVAsO5 | >0.86 | mp-563010 | RbFeMo2O8 | structure not stable |
| mp-6496 | Ba2NaCu3O6 | >0.86 | mp-505042 | CuBi2O4 | structure not stable |
| mp-1191696 | LiCrMo2O8 | >0.86 | mp-19142 | Mn2V2O7 | structure not stable |
| mp-2669 | Mo8O23 | structure not stable | mp-560273 | Cu3WO6 | structure not stable |

**Table S3.** Interstitial oxygen formation energy $E_f$ in bulk $La_4Mn_5Si_4O_{22+\delta}$ under atmosphere environment (T=300 K, $P(O_2)$=0.2 atm), and the optical band gap of $La_4Mn_5Si_4O_{22}$ calculated by different exchange and correlation functionals and from experimental measurement.

| Functional | $E_f$ (eV) | Optical Indirect Band gap (eV) |
|---|---|---|
| GGA | -0.7 | 0.44 |
| GGA+U (U=3.9) | 1.13 | 1.02 |
| SCAN | -0.11 | 0.72 |
| HSE ($\alpha$=0.10) | 0.73 | 0.73 |
| HSE ($\alpha$=0.15) | 1.02 | 1.11 |
| HSE ($\alpha$=0.20) | 1.31 | 1.54 |
| HSE ($\alpha$=0.25) | 1.57 | 1.93 |
| **Experiment** | | 0.79 |

**Table S4.** Comparison of the lattice vectors of $La_4Mn_5Si_4O_{22+\delta}$ obtained from the Rietveld refinement of XRD data, and previous report (PDF: 04-011-1866).[1] The goodness-of-fit quantities are $\chi^2$=2.589, profile R-factor $R_P$= 10.79, and weighted profile R-factor $R_{wP}$= 12.65.

| Parameter | Rietveld refined data | reference data |
|---|---|---|
| a (Å) | 14.0461 (9) Å | 14.024 Å |
| b (Å) | 5.5836 (4)Å | 5.571 Å |
| c (Å) | 11.7299 (8) Å | 11.703 Å |
| $\alpha$ | 90° | 90° |
| $\beta$ | 114.3579 (13)° | 114.34° |
| $\gamma$ | 90° | 90° |

**Table S5.** Rietveld refined lattice positions of $La_4Mn_5Si_4O_{22+\delta}$.

| Elements | Wyckoff position | x | y | z |
|---|---|---|---|---|
| La1 | 4i | 0.2375 (3) | 0 | 0.2596 (3) |
| La2 | 4i | 0.0521 (2) | 0 | 0.7466 (3) |
| Mn1 | 4g | 0 | 0.2585 (9) | 0 |
| Mn2 | 4i | 0.2691 (5) | 0. | -0.0013 (7) |
| Mn3 | 2d | 0 | 0.5 | 0.5 |
| Si1 | 4i | 0.1700 (10) | 0 | 0.5556 (10) |
| Si2 | 4i | 0.4187 (13) | 0 | 0.7505 (17) |
| O1 | 8j | 0.0736 (12) | 0.268 (2) | 0.1769 (16) |
| O2 | 8j | 0.2924 (11) | 0.262 (3) | 0.1160 (13) |
| O3 | 8j | 0.3768 (15) | 0.263 (3) | 0.4068 (15) |
| O4 | 4i | 0.0993 (10) | 0 | 0.9910 (15) |
| O5 | 4i | 0.4125 (20) | 0 | 0.008 (3) |
| O6 | 4i | 0.5062 (15) | 0 | 0.7033 (19) |
| O7 | 4i | 0.2784 (18) | 0 | 0.624 (2) |
| O8 | 4i | 0.1495 (17) | 0 | 0.4060 (20) |

**Table S6.** Atomic percentage value along with the standard deviation measured by the electron probe micro-analyzer (EPMA) and the derived formula from EPMA and Iodometric titration analysis compared with the ideal results.

| | **La** | **Mn** | **Si** | **O** | **Formula** |
|---|---|---|---|---|---|
| Ideal | 11.43% | 14.29% | 11.43% | 62.86% | $La_4Mn_5Si_4O_{22}$ |
| EPMA | 11.39±0.14% | 13.34±0.09% | 11.46±0.29% | 63.81±0.43% | $La_4Mn_{4.69}Si_{4.03}O_{22+0.42}$ |
| Titration | 11.39% (EPMA) | 13.34% (EPMA) | 11.46% (EPMA) | 63.85±0.07 % | $La_4Mn_{4.69}Si_{4.03}O_{22+0.47}$ |

**Supplementary Figures**

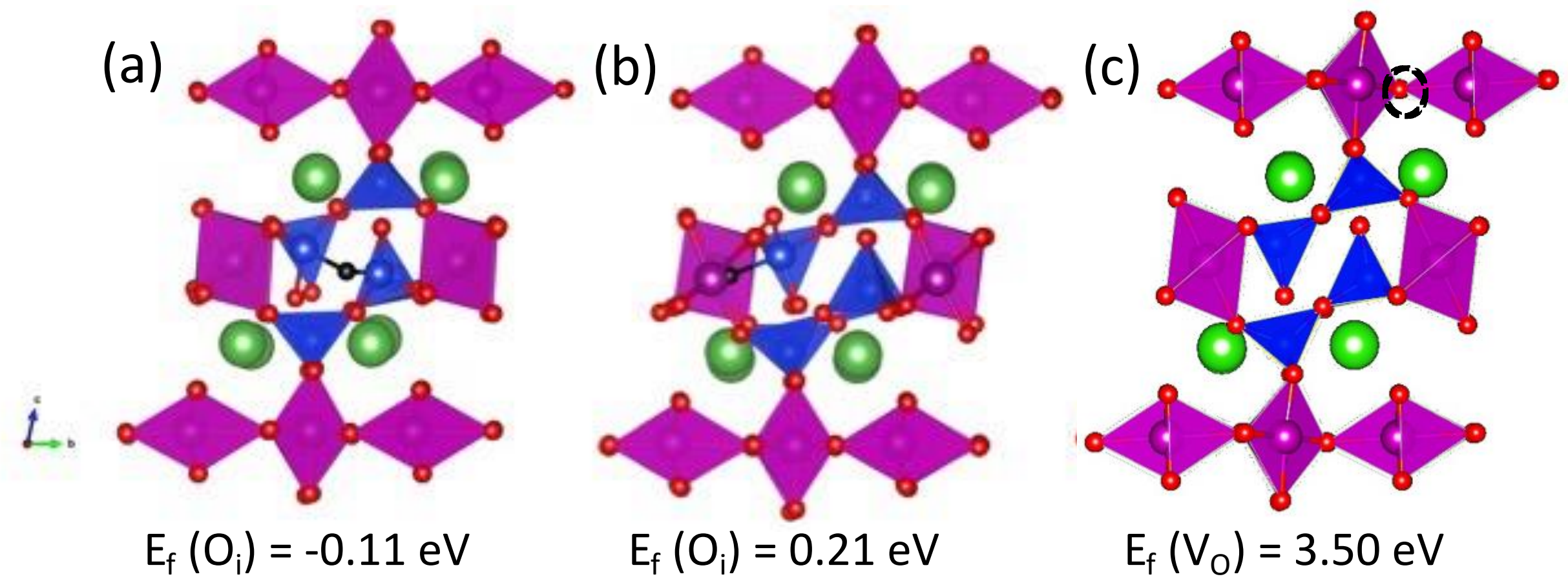

**Figure S1.** Configuration of the two most stable sites of interstitial oxygen $O_i$ marked by black balls in (a) and (b), and the most stable vacancy oxygen $V_O$ site marked by the black dashed circle in (c), along with the defect formation energy under air condition.

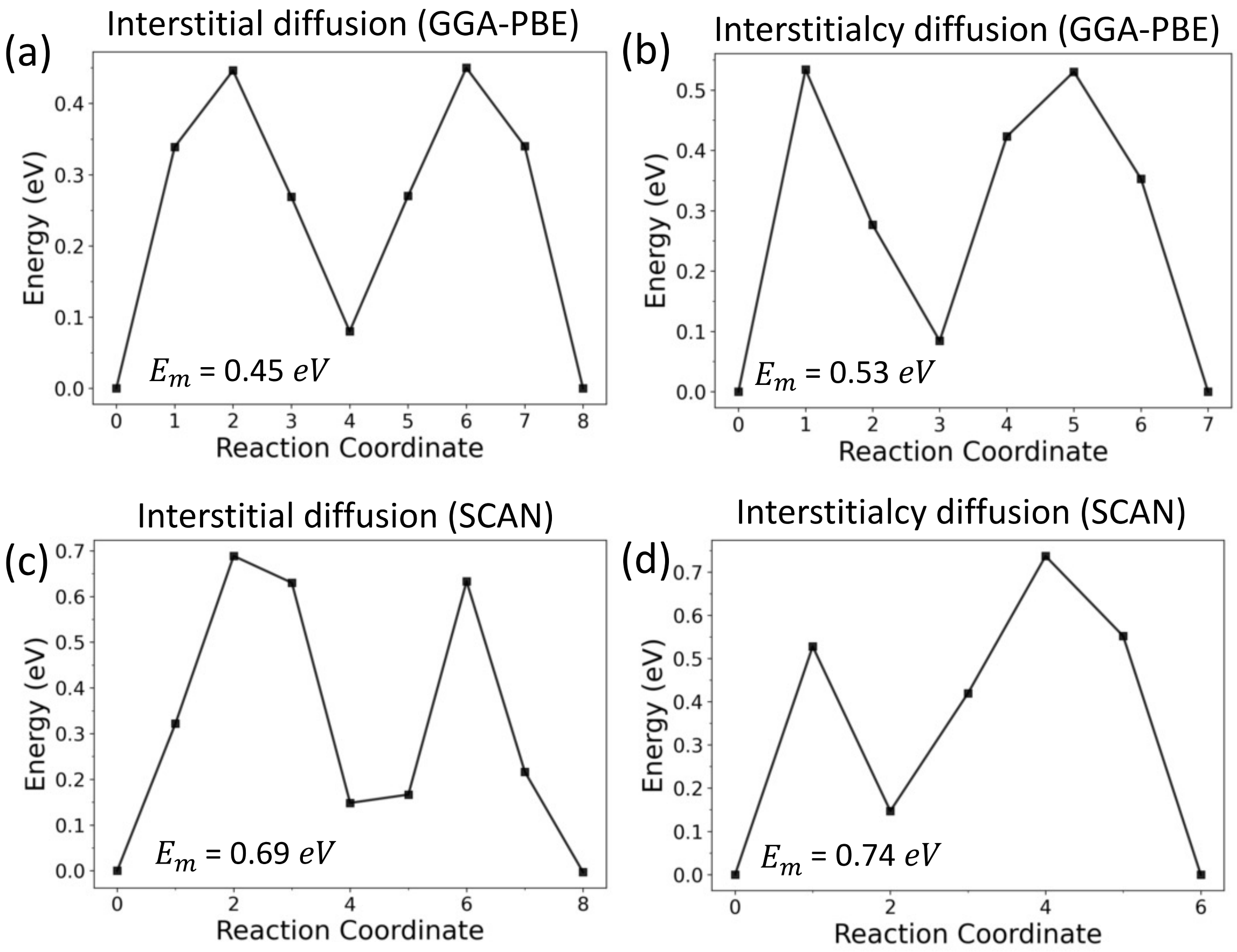

**Figure S2.** Energy landscape of the (a, c) interstitial diffusion and (b, d) interstitialcy diffusion calculated by GGA-PBE, and SCAN functional, respectively.

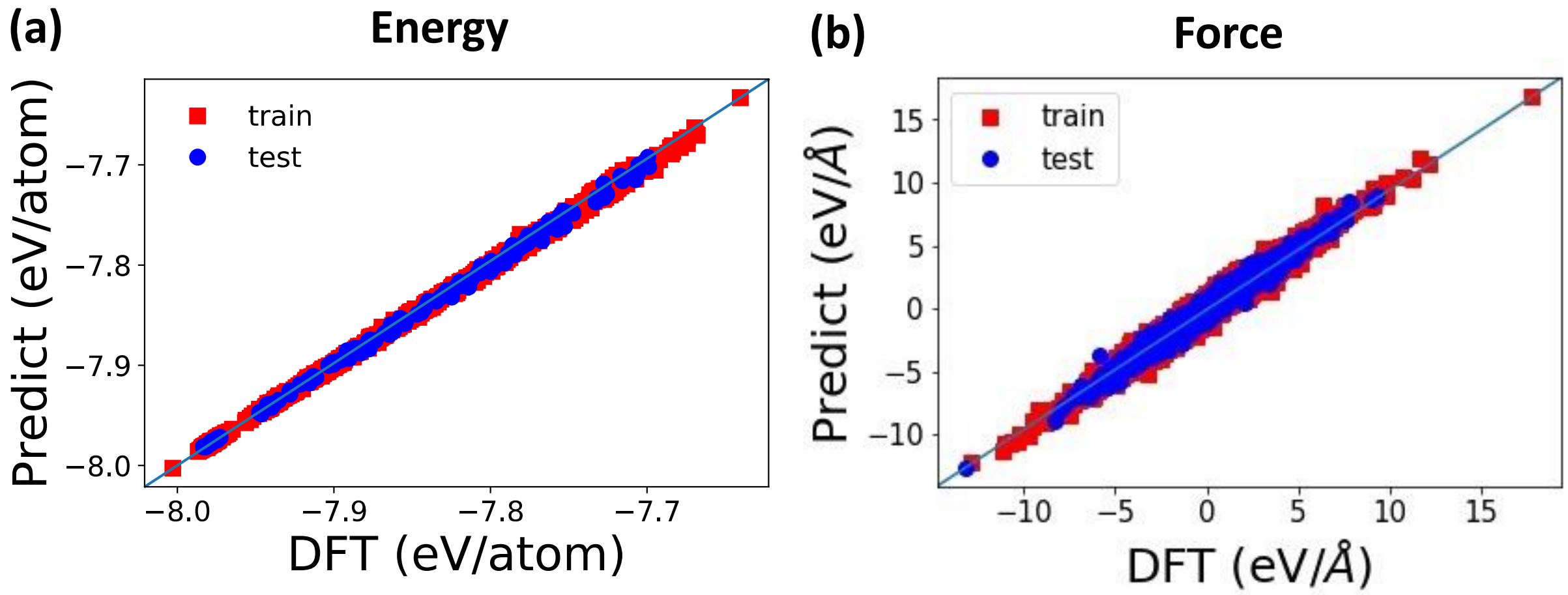

**Figure S3.** Comparison of the machine learning interatomic potential predicted (a) energy, and (b) force on each atom with the DFT results.

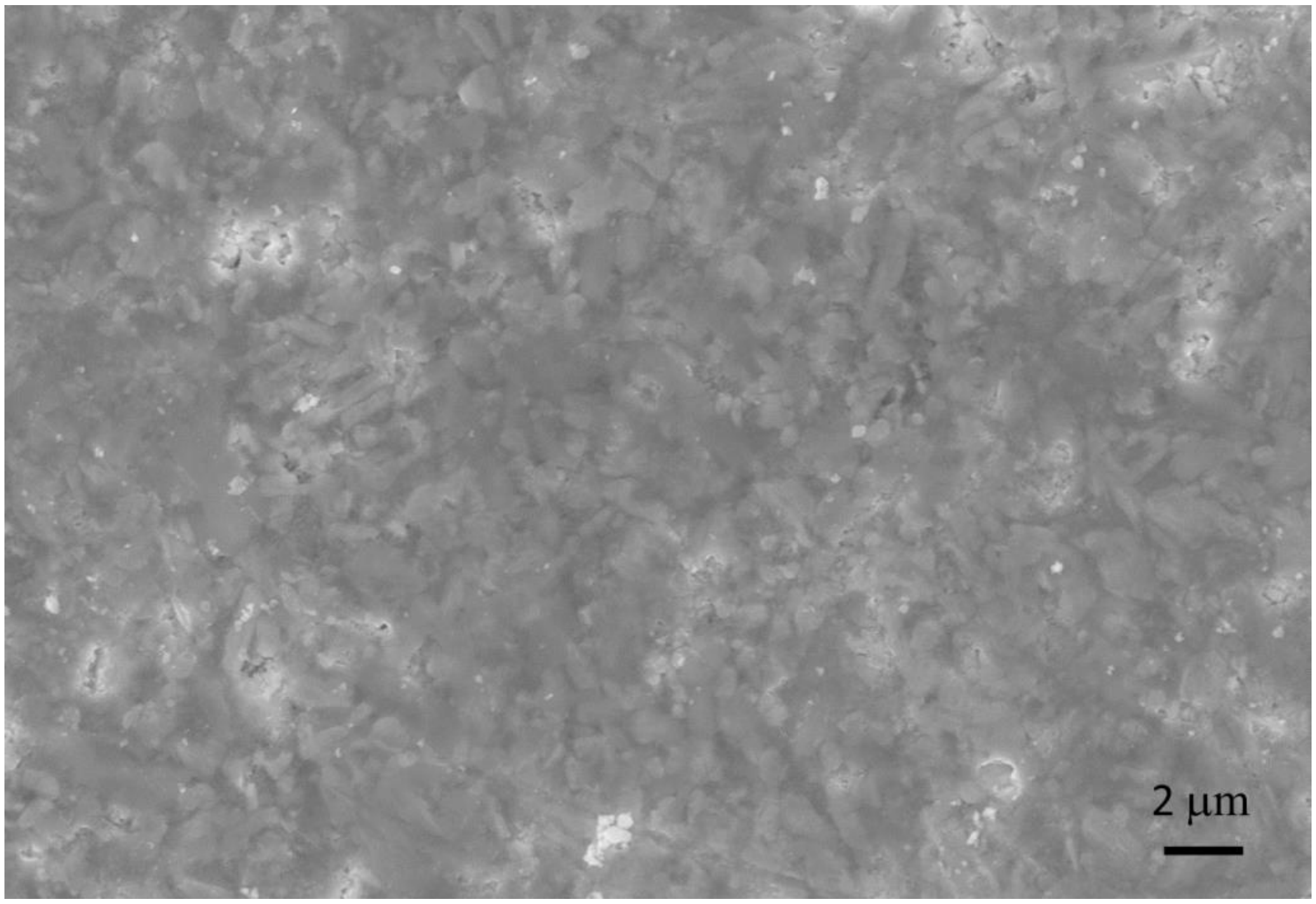

**Figure S4.** The field emission electron microscope (FESEM) image of dense $La_4Mn_{4.69}Si_{4.03}O_{22.42}$ (LMS) pellet sintered at 1050 °C for 24h.

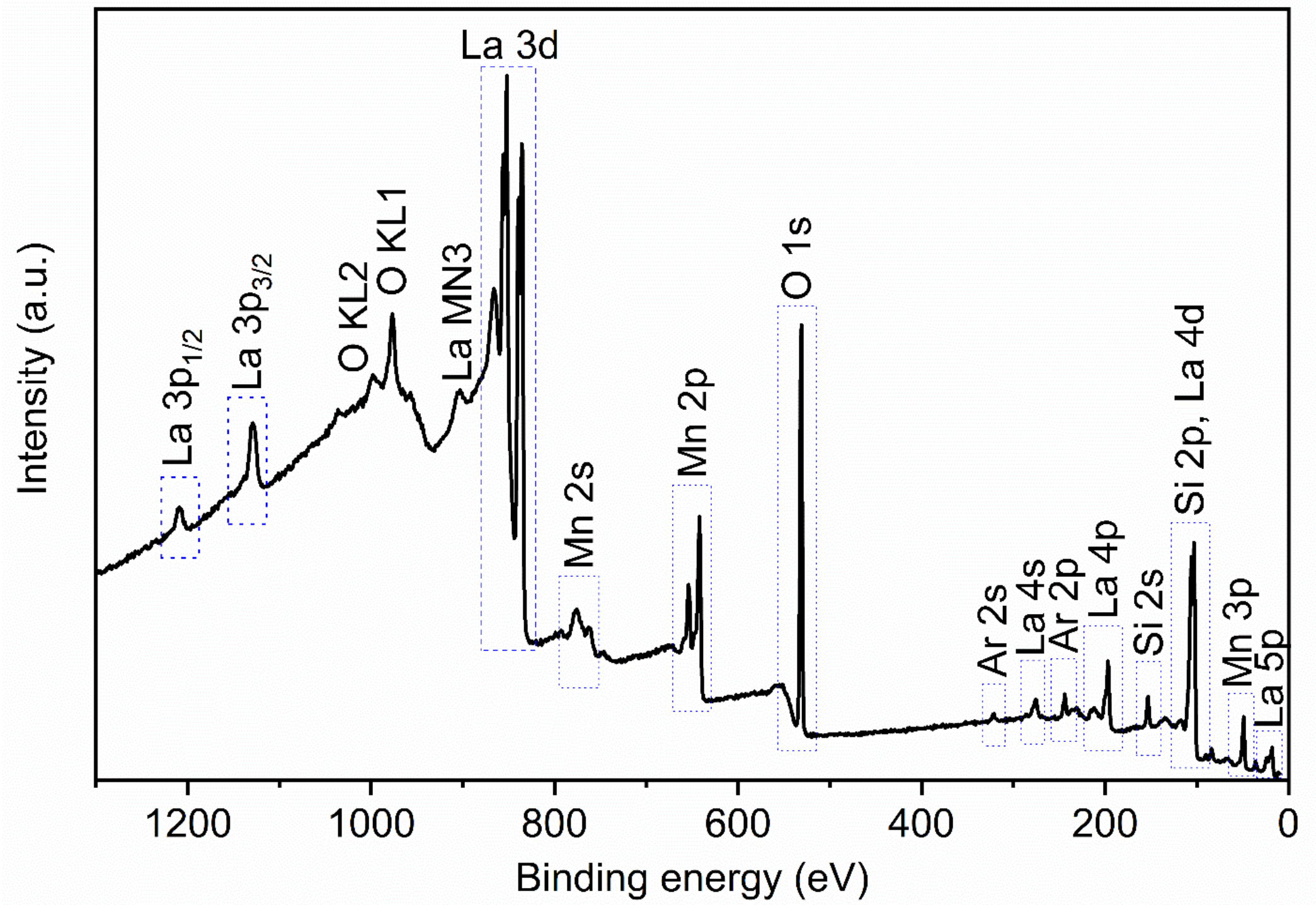

**Figure S5.** XPS survey spectrum of LMS, suggesting the presence of La, Mn, Si, and O atoms without presence of any detectable impurity elements.

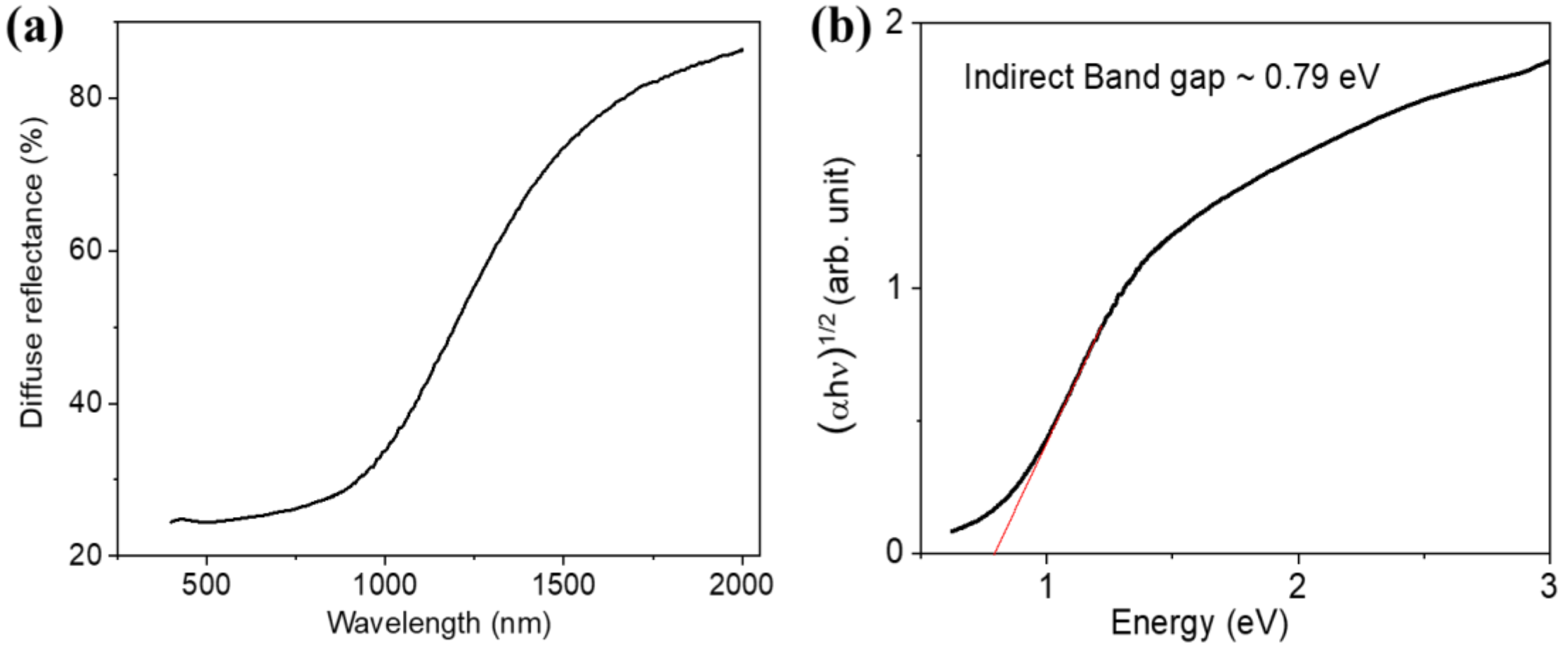

**Figure S6.** (a) Diffuse reflectance spectroscopy of LMS pellet. (b) The optical band gap of LMS was determined from the diffuse reflectance spectroscopy.

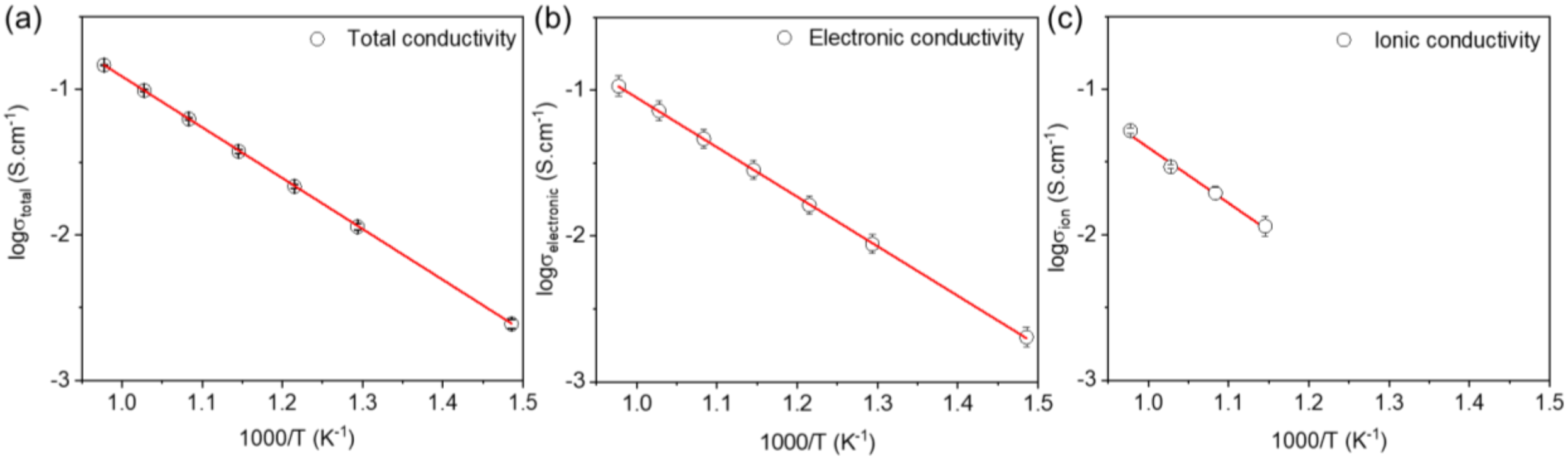

**Figure S7.** Measured (a) total conductivity, (b) electronic conductivity, and (c) ionic conductivity of the $La_4Mn_{4.69}Si_{4.03}O_{22.42}$ sample. The error bars represent the standard deviation. The average conductivity and standard deviation were derived from the conductivity values measured on three different LMS samples. Some error bars are difficult to see as they are less than the size of the symbols.

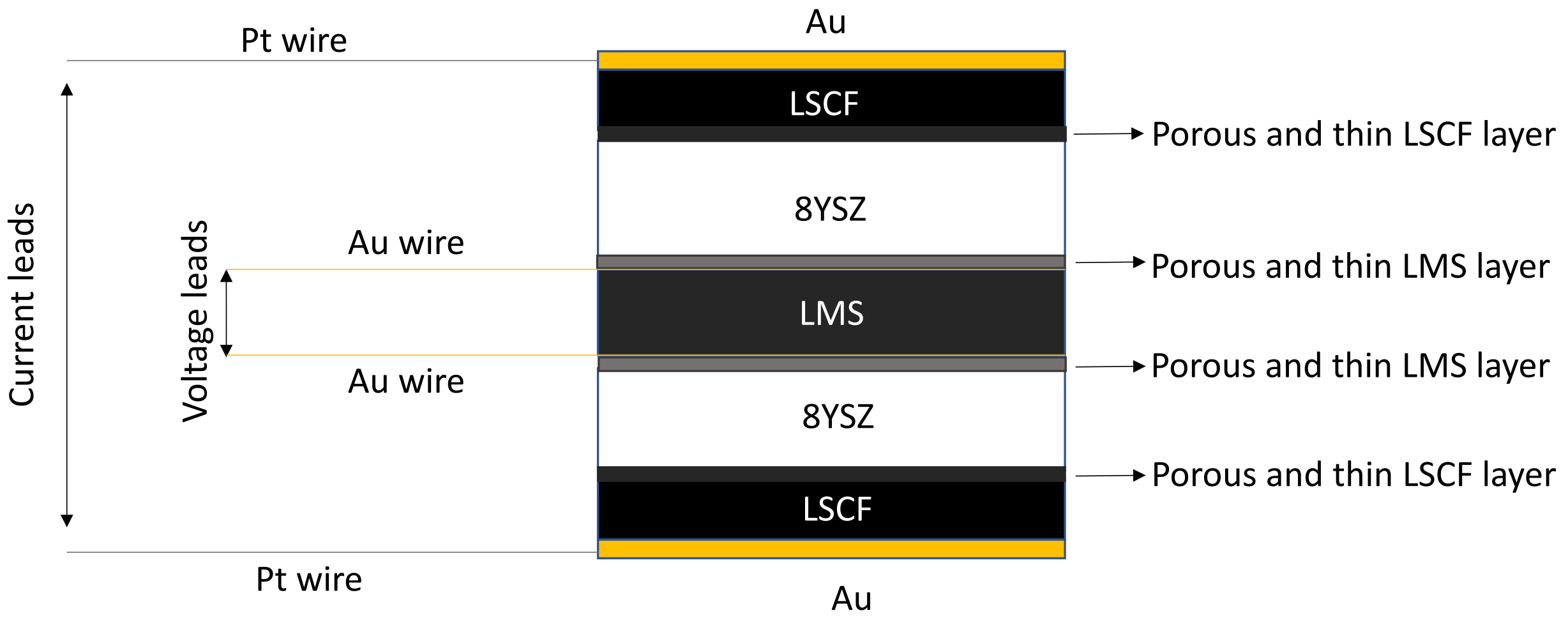

**Figure S8.** A schematic of the ionic conductivity measurement method using the conventional DC 4-probe method with 8YSZ (8 mol% Yttria-Stabilized Zirconia) electron blocking on both ends.

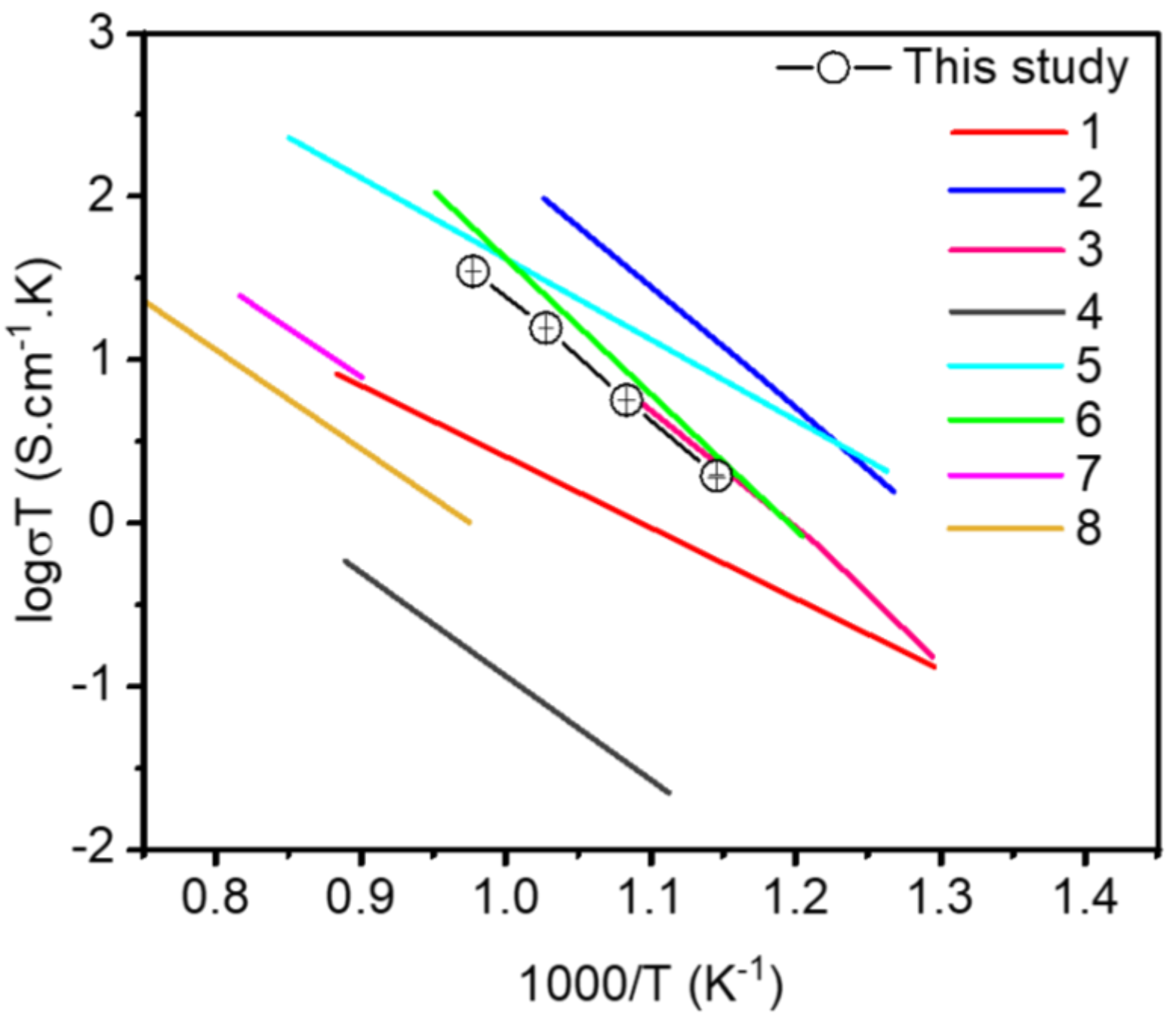

**Figure S9.** Temperature-dependent ionic conductivity of $La_{0.6}Sr_{0.4}Co_{0.2}Fe_{0.8}O_{3-\delta}$ (LSCF) measured using 8YSZ (8 mol% Yttria-Stabilized Zirconia) electron blocking in this work compared with reported ionic conductivity of LSCF measured by impedance spectroscopy (1) $La_{0.6}Sr_{0.4}Co_{0.2}Fe_{0.8}O_{3-\delta}$,[2] (2) $La_{0.54}Sr_{0.44}Co_{0.2}Fe_{0.8}O_{3-\delta}$,[3] and (3) $La_{0.6}Sr_{0.4}Co_{0.2}Fe_{0.8}O_{3-\delta}$;[2] 2-probe method (4) $(La_{0.6}Sr_{0.4})_{0.95}Co_{0.2}Fe_{0.8}O_{3-\delta}$;[4] 4-probe method (5) $La_{0.6}Sr_{0.4}Co_{0.2}Fe_{0.8}O_{3-\delta}$,[5] and (6) $La_{0.8}Sr_{0.2}Co_{0.8}Fe_{0.2}O_{3-\delta}$;[6], and oxygen permeation method (7) $La_{0.65}Sr_{0.3}Co_{0.2}Fe_{0.8}O_{3-\delta}$,[7] and (8) $La_{0.8}Sr_{0.2}Co_{0.1}Fe_{0.9}O_{3-\delta}$.[7] Voltage measurement at a fixed temperature on one LSCF sample was performed at a regular interval of 0.259 s for 10 min. Nearly 2.3k data points were averaged for the conductivity determination at a temperature. The error bars represent the standard deviation derived from these nearly 2.3k data points collected at a temperature. The error bars are difficult to see as they are smaller than the size of the symbols.

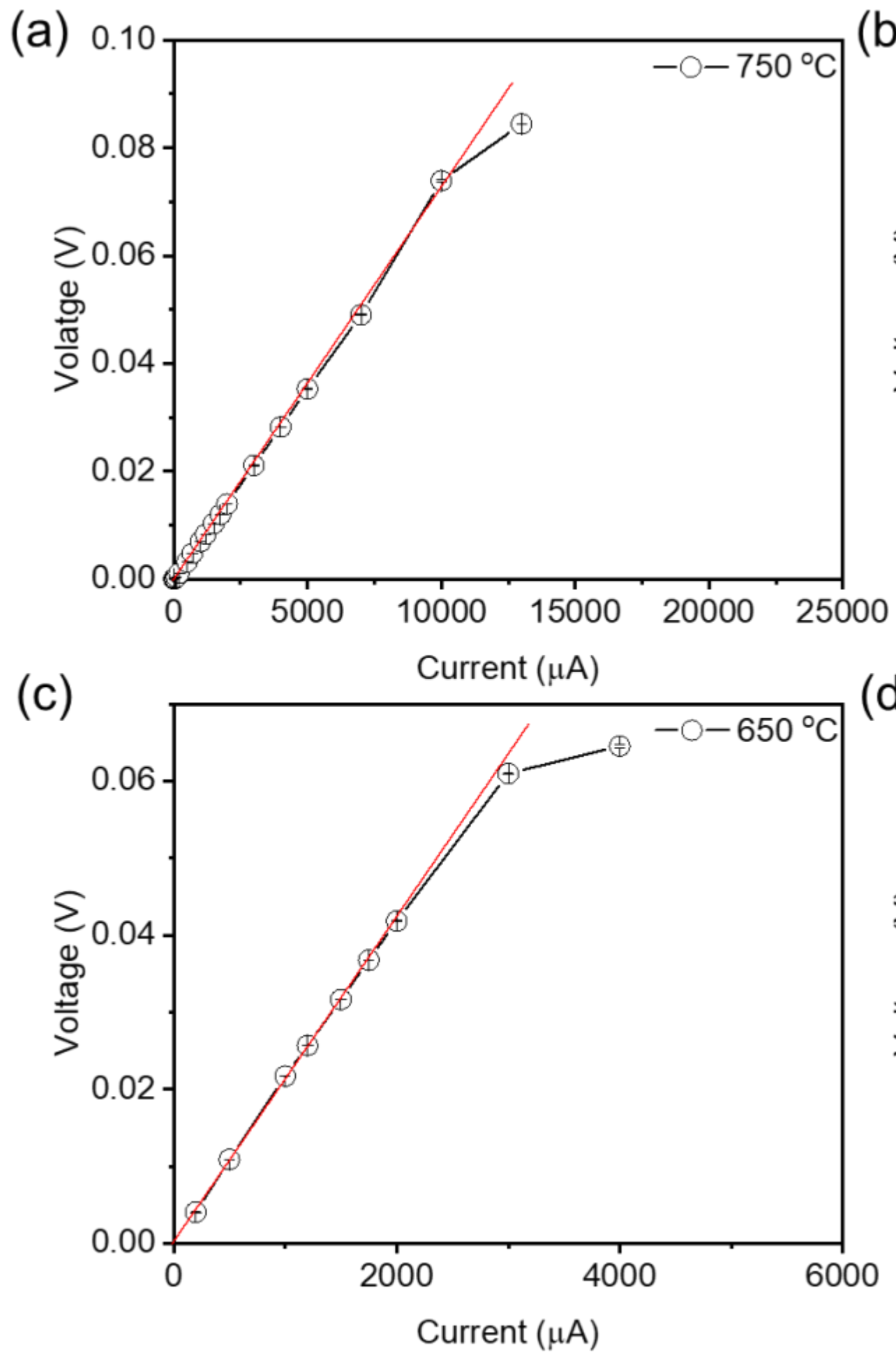

**Figure S10.** Current-voltage characteristics across LMS in one Au/LSCF/YSZ/LMS/LSCF/YSZ/Au system at (a) 750 °C, (b) 700 °C, (c) 650 °C, and (d) 600 °C. The error bars represent the standard deviation obtained from ~150 measurements at each fixed current from one sample. The error bars are difficult to see as they are less than the size of the symbols.

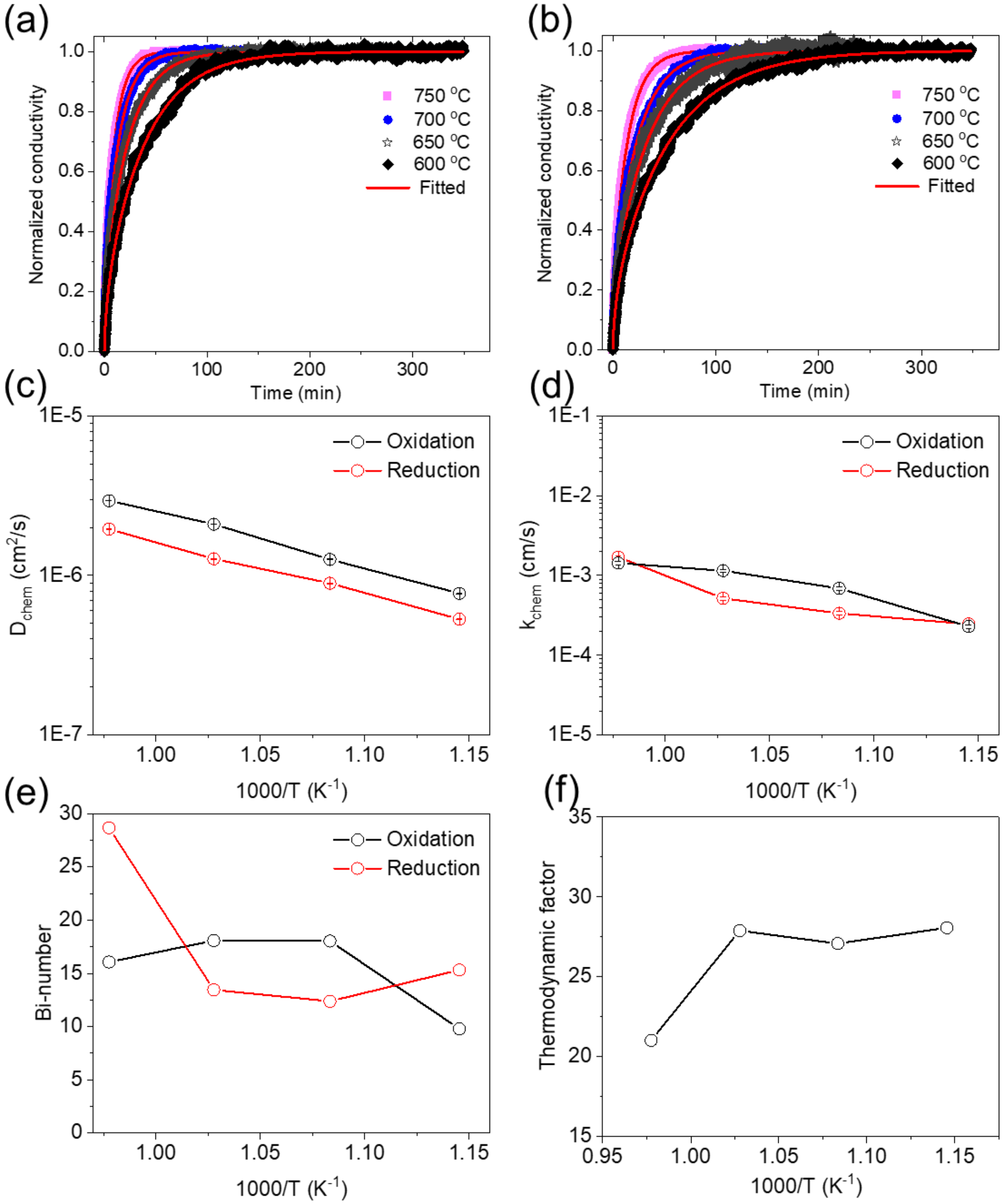

**Figure S11.** Normalized and fitted electrical conductivity relaxation (ECR) data of one LMS sample during the (a) oxidation and (b) reduction at different temperatures. The fitted (c) $D_{chem}$, (d) $k_{chem}$ and (e) calculated Bi-number during oxidation and reduction. (f) Calculated thermodynamic factor at the studied temperatures. The error bars of the fitted values (with 95% confidence, i.e. ±2 standard deviation) are smaller than the size of the symbols. Please see details of fitting of the ECR data in Discussion 7.

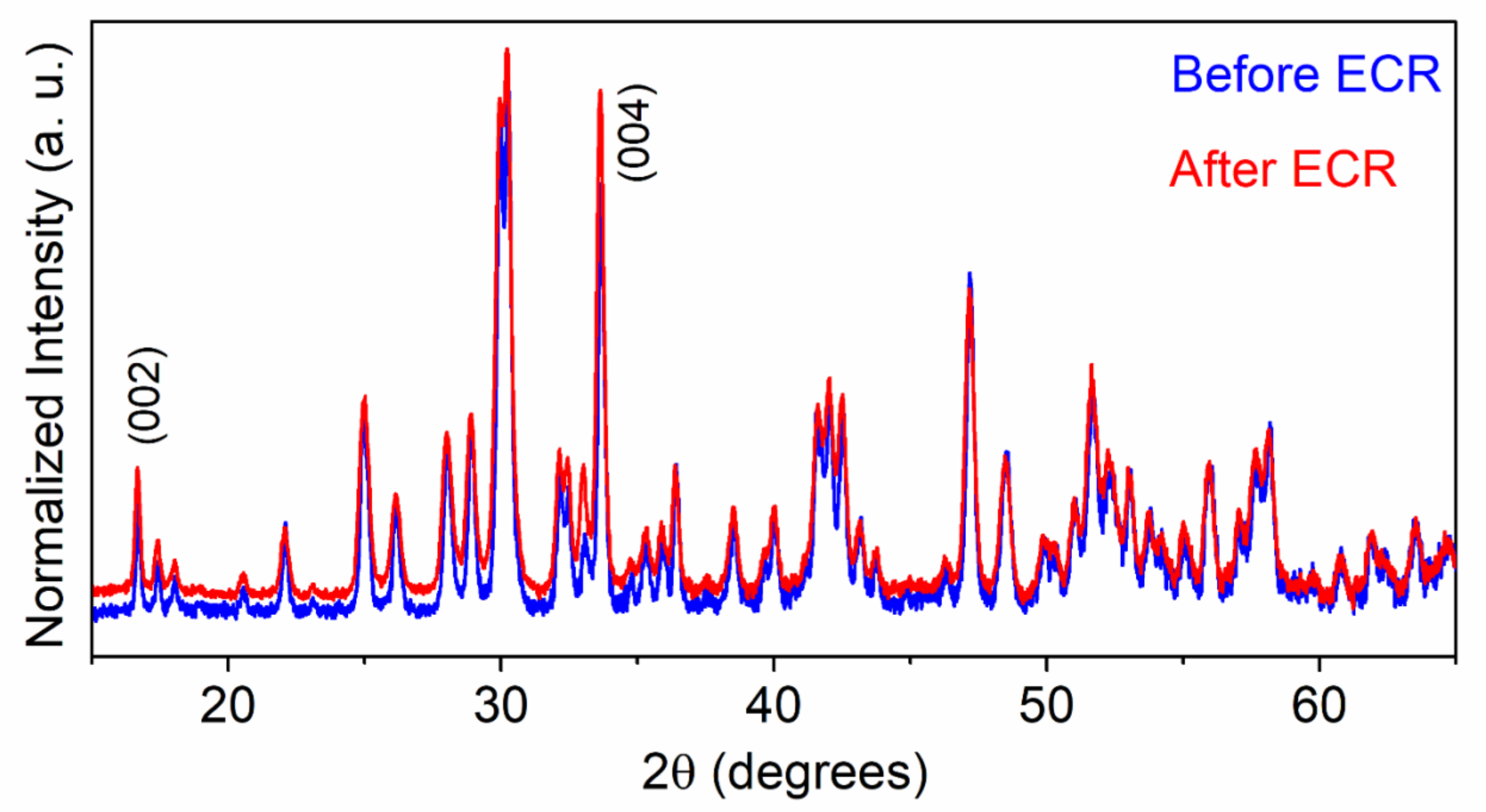

**Figure S12.** Room temperature X-ray diffraction (XRD) pattern of the LMS pellet before and after the electronic conductivity relaxation (ECR) study confirms the structural stability. The XRD profile also demonstrates enhanced grain orientation on the pellet surface along the (001) direction.

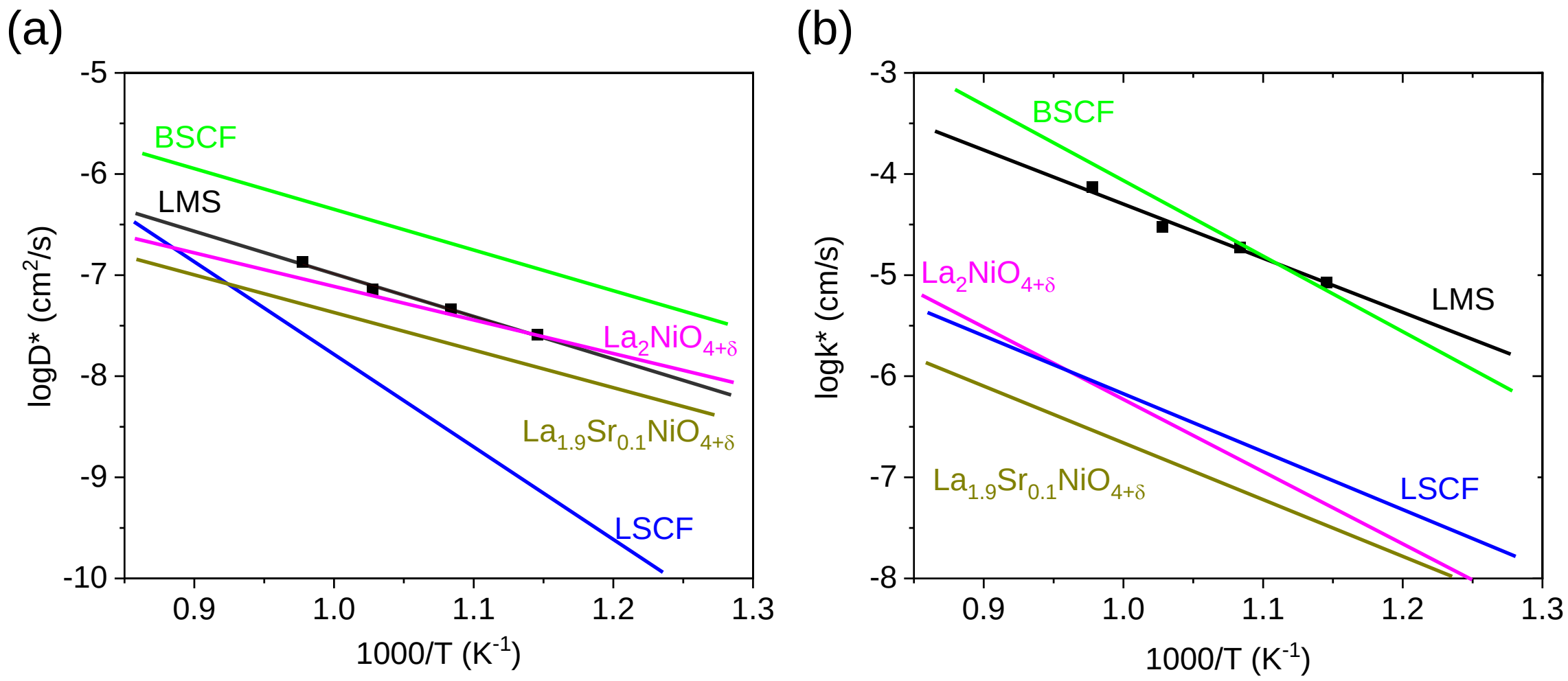

**Figure S13.** (a) The tracer diffusion coefficient $D^*$ of $La_4Mn_5Si_4O_{22+\delta}$ compared with $La_{0.6}Sr_{0.4}Co_{0.2}Fe_{0.8}O_3$ (LSCF),[8–10] $Ba_{0.5}Sr_{0.5}Co_{0.8}Fe_{0.2}O_3$ (BSCF),[11,12] $La_2NiO_{4+\delta}$,[13–16] and $La_{1.9}Sr_{0.1}NiO_{4+\delta}$.[13,15] (b) The trace surface exchange coefficient $k^*$ of $La_4Mn_5Si_4O_{22+\delta}$ compared with $La_{0.6}Sr_{0.4}Co_{0.2}Fe_{0.8}O_3$ (LSCF),[8,17] $Ba_{0.5}Sr_{0.5}Co_{0.8}Fe_{0.2}O_3$ (BSCF), [11,12] $La_2NiO_{4+\delta}$,[13–15,18] and $La_{1.9}Sr_{0.1}NiO_{4+\delta}$. [13,15]

## Discussion

### 1. Electrical property and defect formation energy with different DFT functionals.

Since the GGA functional predicts underestimated oxygen ion migration barrier and optical band gap compared to the experiments, the band gap and defect formation energy for the bulk structure of $La_4Mn_5Si_4O_{22+\delta}$ were studied with different DFT exchange and correlation functionals, which are the GGA, GGA with Hubbard U correction (GGA+U) (U=3.9eV for Mn),[19] hybrid functional of Heyd, Scuseria, and Ernzerhof (HSE06),[20] and the strongly constrained and appropriately normed (SCAN) functionals.[21] Formation energy and the band gap vary with different potentials, from which we believe that different approaches to represent the d-electron physics have a significant effect on the defect formation energies. Based on the previous studies on the performance of different exchange-correlation potentials, SCAN predicts the most accurate optical band gap and defect formation energy that is consistent with the experimental observation.

### 2. DFT studies of oxygen defects in $La_4Mn_5Si_4O_{22+\delta}$.

The formation energy of the oxygen interstitial and vacancy (at ≈2% concentration) in the bulk structure of $La_4Mn_5Si_4O_{22}$ were calculated under atmospheric conditions (T=300 K, P($O_2$)=0.2 atm). The two most energetically favorable configurations of the interstitial oxygen site were displayed in **Fig. S1a-b** and the most stable configuration of the vacancy oxygen site was shown in **Fig. S1c**, along with the defection formation energies. These calculations were performed using the SCAN functional,[22] which was proven to give the most consistent band gap value with experiment (**Discussion 1**). Monkhorst−Pack k-point meshes[23] of 4 × 3 × 2 was used for the 2 × 1 × 1 supercell with 70 atoms.

The formation energies of $O_i$ at different concentrations in $La_4Mn_5Si_4O_{22+\delta}$ were studied, which are -0.11 eV, -0.11 eV, 0.49 eV, and 0.52 eV under air conditions for $\delta = 0.25, 0.5, 0.75, 1$, respectively. The results suggest that the interstitial is stabilized by oxidizing $Mn^{2+}$ ions. With including one interstitial oxygen into the 2 × 1 × 1 supercell ($\delta = 0.5$), the only two $Mn^{2+}$ ions in the supercell were oxidized to $Mn^{3+}$. With more interstitial oxygen included ($\delta = 0.75$), the defect formation energy is increased by 0.6 eV, indicating that further oxidization of $Mn^{3+}$ ions is difficult. It is worth noting that the defect formation energies are very close when $\delta = 0.25, 0.5$ and $\delta = 0.75, 1$, respectively, suggesting that the interstitial concentration dependence of the formation energy is much more strongly impacted by the oxidative state of Mn ions than the direct interstitial interaction. The results suggest that the equilibrium concentration of interstitial oxygen in $La_4Mn_5Si_4O_{22+\delta}$ is $\delta \approx 0.5$ under air conditions, which is consistent with the EMPA measurement of the interstitial content $\delta = 0.42$ and the TGA results of $\delta = 0.42 \sim 0.52$.

### 3. Calculated spin state of $La_4Mn_5Si_4O_{22+\delta}$.

The DFT calculated magnetic moments are consistent with $Mn_1^{4+}$, $Mn_2^{3+}$ and $Mn_3^{2+}$ being in their high spin magnetic states. More specifically, our average integrated (within the Wigner-Seitz radius) z-component of the spin on $Mn_1$, $Mn_2$, and $Mn_3$ are 2.90, 3.41, and 4.41 $\mu_B$ from DFT GGA-PBE calculation, respectively. These values are consistent with the expected ideal formal valence spin state based on summing of unpaired electrons of 3, 4, and 5 for $Mn_1^{4+}$, $Mn_2^{3+}$ and $Mn_3^{2+}$, respectively. These ideal formal valence spin values can be used to calculate the total spin-only magnetic moment from the formula $\mu_{cal}^2 = 2 * \mu_{cal}^2(Mn^{4+}) + 2 * \mu_{cal}^2(Mn^{3+}) + \mu_{cal}^2(Mn^{2+})$ /formula unit,[1] where for each Mn ion the effective magnetic moment was calculated by $\mu_{cal} = \sqrt{n * (n + 2)}$ and $n$ is the sum of unpaired electron. Using this formula, the total spin-only magnetic moment is predicted to be 10.63 $\mu_B$/formula unit, which is an excellent match for the experimental measured value of 10.64 $\mu_B$/formula.[1] Upon the inclusion of $O_i$, two adjacent $Mn^{2+}$ ions are oxidized to $Mn^{3+}$, with a change of the spin state from 5 to 4.

**4. Climbing Image Nudged Elastic Band (CI-NEB) calculation.**

The migration barriers of the $O_i$ interstitial and interstitialcy diffusion pathways in LMS were studied by the Climbing Image Nudged Elastic Band (CI-NEB) method.[24] The calculations were performed separately using GGA-PBE and SCAN functionals. The plane wave cutoff energy was set as 520 eV. The stopping criteria for total energy calculations were 0.001 meV/cell for electronic relaxation and 0.05 eV/Å for ionic relaxation, respectively. 7 images were used for the interstitial diffusion pathway and 5 images were used for the interstitialcy diffusion pathway.

**5. Iodometric titration.**

Iodometric titration was performed in nitrogen atmosphere based on the following assumptions/criteria: First, charge neutrality; second, cations are in the stoichiometric ratio as obtained in the EPMA study; third, the valence of La and Si in LMS is 4+; fourth, average valence of Mn in LMS is X+ > 2+. The reaction mechanism is

$$Mn^{X+} + (X-2)Cl^- = Mn^{2+} + (X-2)/2.\ Cl_2$$

$$Cl_2 + 2I^- = 2Cl^- + I_2$$

A weighed amount of ground LMS pellet was dissolved in an aqueous solution of KI and HCl (6N). $Cl_2$ is generated during this dissolution and the in-situ generated $Cl_2$ reacts with the I- to produce $I_2$. The liberated $I_2$ is measured by titration with a standard volumetric aqueous solution of $Na_2S_2O_3$ (~ 0.01 N). Finally, the stoichiometry of the oxygen was calculated from the measured $I_2$ amount. The measurement was repeated for five times to confirm the reproducibility, and the average value along with the standard deviation in the mean was presented in **Table S6**.

**6. Preparation of the ionic conductivity measurement using YSZ blocks.**

The ionic conductivity measurement was performed using pre-synthesized commercial electrode material yttria-stabilized zirconia (8 mol % $Y_2O_3$ in ZrO2, 8YSZ; Sigma Aldrich) blocks as the electronic blocking electrode.[4] The 8YSZ pellets were sintered at 1500 °C for 6 hours. As shown in Fig. S8, the cross-section of the 8YSZ pellets was 4.8 mm x 4.8 mm and the thickness was 1.5 mm. The thickness of the LMS pellets were ~ 0.8 mm. All pellets were polished on all sides to remove any surface contamination and to reduce contact resistance. To ensure better connectivity between LMS and 8YSZ pellet we used a homemade paste of LMS in ethanol and made a thin layer of LMS between LMS and 8YSZ blocks. For voltage measurement, two thin Au wires were inserted at the YSZ and LMS junctions for the voltage measurement. To measure the voltage across the LMS pellet accurately, we sputtered an Au line (width ~ 0.2 mm, thickness~ 200 nm) on the LMS surface and connected it with the Au wires. For the efficient oxygen exchange, we used two porous LSCF pellets (thickness ~ 1 mm) at two ends of the assembly. The porous LSCF pellets were sintered at 1050 °C for 6 hours using commercial LSCF electrode power (Sigma Aldrich). These two sintered porous LSCF pellets were also connected to the YSZ pellet by a thin LSCF layer made using homemade LSCF paste. The exposed surface of the LSCF pellets was coated with Au by sputtering. The whole sample assembly was pressed vertically between two Au-coated alumina plates and two Pt wires were connected to these alumina plates as the current leads. The whole system was sintered at 900 °C for 1 hour before performing the measurement.

**7. Fitting of the ECR data**

The obtained ECR data was fitted to determine the $D_{chem}$, $k_{chem}$ along with its standard deviation in the average value, using a previous reported 2-D model.[25] In **Fig. S11(a, b),** the non-linear least square fitting was performed using the publicly available NETL Electrical Conductivity Relaxation (ECR) Analysis Tool. [25] A small conductivity drift was observed during the whole ECR process which may happen due to the microstructure change occurring at high temperatures. We corrected this resistance drift before fitting. The ECR data shown a very fast oxygen exchange at the beginning (~30 s to 2 min, with the time being inversely proportional to temperature) of the transient response, which may be due to the presence of (001) oriented grain on the pellet surface (confirmed by XRD, **Fig. S12**) or a patch of unknown secondary surface phase sufficiently thin as to not be detectable by XRD. In order to fit the ECR data using a single phase model, we neglected this fast response region as fitting this region using the single phase model gives a very high value of $k_{chem}$ with a large error bar. We have chosen the initial point where the percentage of error in both $D_{chem}$ and $k_{chem}$ is < 5%. The error bars reported for the $D_{chem}$ and $k_{chem}$ represent a 95% confidence interval for $D_{chem}$ and $k_{chem}$ that is provided as part of the NETL ECR Analysis Tool[25] based on numerical aspects of their fitting. However, it is important to acknowledge that the true uncertainty in the values of $D_{chem}$ and $k_{chem}$ can be affected by many issues, including sample to sample variability, $D_{chem}$ and $k_{chem}$ covariance in fitting [25] and multiple measurement limitations, e.g., gas flush effects.[26] Variations in $D_{chem}$ and $k_{chem}$ between different samples, research groups, and experimental setups and approaces are often close to an order of magnitude, even for well-studied materials. The value of $D_{chem}$ and $k_{chem}$ of LMS at a fixed temperature was determined as the average during oxidation and reduction at that temperature. The potential for reliable determination of both $D_{chem}$ and $k_{chem}$ was confirmed by the Bi-number, defined as $Bi = \frac{L_Z}{D_{chem}/k_{chem}}$ , where $L_Z$ represents the half-thickness of the LMS sample. The obtained values of Bi number falls within the range of 0.03 to 30[27] (**Fig. S11e**), suggesting that reliable values of both $D_{chem}$ and $k_{chem}$ can be extracted from the measurement. The stability of the samples after the ECR measurement was confirmed using the XRD technique (**Fig. S12**).

**8. Determination of average $D_{chem}$ and $k_{chem}$ of state-of-art materials from literatures.**

Diffusion coefficient $D$ and surface exchange coefficient $k$ of $La_4Mn_5Si_4O_{22+\delta}$ were compared with state-of-the-art materials $La_{0.6}Sr_{0.4}Co_{0.2}Fe_{0.8}O_8O_{3-\delta}$ (LSCF), $Ba_{0.5}Sr_{0.5}Co_{0.8}Fe_{0.2}O_{3-\delta}$ (BSCF),$La_{0.5}Sr_{0.5}CoO_{3-\delta}$ (LSC), $La_{0.5}Sr_{0.5}FeO_{3-\delta}$ (LSF), $BaCoFeYO_{3-\delta}$ (BCFY), and $La_2NiO_{4+\delta}$ (**Fig. 4(e,f) and Fig. S13**). Data along with references are available as digital SI. and We found that there is a wide variation of the $D$ and $k$ reported at different temperatures, making the comparison difficult. Hence, we have calculated the average of $D$ and $k$ at 600 °C and 800 °C from the literatures where the authors have studied at least two different temperatures. From the average $D$ and $k$, we determined the activation energy using the Arrhenius relationship and derived $D$ and $k$ changing with temperature.